\documentclass[traditabstract]{aa}
\usepackage{graphicx,natbib}
\usepackage{txfonts,hyperref}

\def\lta{ \lower .75ex\hbox{$\sim$} \llap{\raise .27ex \hbox{$<$}} }

\begin{document}

\date{Received .../Accepted ...}

\title{Expected disk wind properties evolution along an X-ray Binary outburst}
\author{P.-O. Petrucci\inst{1}
  \and S. Bianchi\inst{2}
  \and G. Ponti\inst{3}
  \and J. Ferreira\inst{1}
  \and G. Marcel\inst{4}
 \and F. Cangemi\inst{5}
  \and S. Chakravorty\inst{6}
  \and M. Clavel\inst{1}
  \and J. Malzac\inst{7}
  \and J. Rodriguez\inst{5}
  \and S. Barnier\inst{1}
  \and R. Belmont\inst{5}
  \and S. Corbel\inst{5}
  \and M. Coriat\inst{7}
  \and G. Henri\inst{1}
  }
  
\institute{Univ. Grenoble Alpes, CNRS, IPAG, F-38000 Grenoble, France 
	   \and
	   Dipartimento di Matematica e Fisica, Universit\`a degli Studi Roma Tre, 
	   via della Vasca Navale 84, 00146 Roma, Italy 
	   \and
	   INAF-Osservatorio Astronomico di Brera, Via Bianchi 46, 23807 Merate, LC, Italy
	   \and
	Villanova University, Department of Physics, Villanova, PA 19085, USA
	\and
	   AIM, CEA, CNRS, Universit\'e Paris-Saclay, Universit\'e de Paris, F-91191 Gif-sur-Yvette, France 
	   \and
	   Department of Physics, Indian Institute of Science, Bangalore 560012, India
	   \and
	   IRAP, Universit\'e de toulouse, CNRS, UPS, CNES , Toulouse, France
} 

%

\abstract{Blueshifted X-ray absorption lines (preferentially from Fe\ XXV and Fe\ XXVI present in the 6--8 keV range) indicating the presence of massive hot disk winds in Black Hole (BH) X-ray binaries (XrB) are most generally observed during the soft states. It has been recently suggested that the non-detection of such hot wind signatures in the hard states could be due to the thermal instability of the wind in the ionization domain consistent with Fe\ XXV and Fe\ XXVI. Studying the wind thermal stability requires however a very good knowledge of the spectral shape of the ionizing Spectral Energy Distribution (SED). We discuss in this paper the expected evolution of the disk wind properties during an entire outburst by using the {\it RXTE} observations of GX 339-4 during its 2010--2011 outburst. While GX 339-4 never showed signatures of a hot wind in the X-rays, the dataset used is optimal to illustrate our purposes. We compute the corresponding stability curves of the wind using the SED obtained with the Jet-Emitting Disk model. We show that the disk wind can transit from stable to unstable states for Fe\ XXV and Fe\ XXVI ions on a day time scale. While the absence of wind absorption features in hard states could be explained by this instability, their presence in soft states seems to require changes of the wind properties (e.g. density) during the spectral transitions between hard and soft states. We propose that these changes could be partly due to the variation of heating power release at the accretion disk surface through irradiation by the central X-ray source. The evolution of the disk wind properties discussed in this paper could be confirmed through the daily monitoring of the spectral transition of a high-inclination BH XrB.
%
%
}


\maketitle

\section{Introduction}
Low mass X-ray binaries (XrB) are binary systems containing a main sequence star and a compact object (neutron star or black hole). Due to Roche lobe overflow, the matter is pulled off from the star and forms an accretion disk around the compact object (see \citealt{tau06} for a review). XrBs spend most of their time in a quiescent state at very low mass accretion rates. Occasionally, they come out of the quiescent state and undergo outbursts that last from a few months to a year, during which their flux rises by several orders of magnitude across the whole electromagnetic spectrum (e.g. \citealt{rem06}). This release of gravitational power is believed to result from disk instabilities in the outer part of the accretion flow, driven by the ionization of hydrogen above a critical temperature (e.g. \citealt{mey81, sma84} and \citealt{ham19} for a recent review).\\

During the outburst, XrBs {{exhibit a large panel of spectral and timing properties}}. Black hole (BH) XrB (which are the one we are interested in this paper and labeled BH XrB or simply XrB in the following) can be easily distinguished in the so-called Hardness-Intensity Diagram (HID) where the X-ray luminosity is plotted against the hardness ratio of the X-ray spectrum, producing a hysteresis with a typical Q-shaped track. {{All BH XrBs generally follow the same spectral evolution}} (e.g. \citealt{hom05, dun10}, and see \citealt{rem06} or \citealt{don07} for reviews). At the beginning and end of the outburst, the system is in the so-called hard state. During this state, the X-ray spectrum has a hard power-law shape up to few tens of keV, signature of non-thermal processes in a very hot, optically thin, plasma (the so-called corona). The radio emission observed in this state, that extends up to the Infrared (IR), is believed to be entirely produced by a steady jet. 
During the middle part of the outburst, the system is in the so-called soft state where the spectrum is dominated by a bump in the soft X-rays ($<$ 2 keV) commonly interpreted as the thermal emission from the inner region (close to the innermost stable circular orbit, hereafter ISCO) of an optically thick accretion disk. In this state, the radio/IR emission is strongly reduced (e.g. \citealt{cori09}) and even undetectable in most cases, suggesting the disappearance (or strong fading) of the jet component. In between these two main states, the system transitions, in a few days, through the so-called (hard and soft) intermediate states. The spectral shape evolves from hard-to-soft, in the first part of the outburst, and then from soft-to-hard, at the end of the outburst before turning back to the quiescent state (e.g. \citealt{rem06,don07}).\\

In the last 15 years it has been realized that, while soft states do not show signatures of jet, the presence of blue shifted absorption lines, generally from Fe XXV and Fe XXVI, indicates the existence of massive hot disk winds (see \cite{dia16} and \cite{pon16} for recent reviews). While {{not detected}} in the hard states, these winds are generally present in the soft states of inclined systems (e.g. \citealt{pon12}). But the picture is certainly more complicated than a simple on-off wind process between spectral states (e.g. \citealt{mun16,hom16,san20}), and  recent results during the (hard and soft) intermediate states show a large variety of wind properties and signatures  \citep{mil08,kal09,rah14,nei16,shi16,mun17,mat18,gat19}. The physical processes at the origin of these disk winds are not completely understood but observations favored thermal or magnetic driving (or a combination of the two) as the most probable launching mechanisms (e.g. \citealt{dia16,tet18a}). 
The dependence of the wind detection with the spectral state of the X-ray source is however not straightforward. Indeed, assuming that the outflow velocity is of the order of the escape velocity of the absorbing material, the observed blue-shifts generally put the absorbing wind at up to few/hundreds of thousand of gravitational radii away from the X-ray emitting region located close to the black hole \citep{schu02,ued04,kub07,mil08b,kal09}. In XrBs, such distances correspond to the outer part of the accretion disk. This connection between the inner and outer part of the accretion flow cannot occur on accretion (i.e. viscous) timescale since it is expected to be of the order of several weeks to months, i.e. the typical duration of the observed outbursts\footnote{For an $\alpha$-disk \citep{sha73} around a 10 solar-mass black hole characterized by a disk aspect ratio $H/R$, with $H$ the height of the disk at the radius $R=rR_g$, the accretion timescale to reach the black hole from the disk radius $R$ is $t_{visc}= 5\times 10^{-5} (H/R)^{-2}\alpha^{-1}r^{3/2}$ seconds. For $H/R=10^{-2}$, $\alpha=0.1$ and $R=10^4 R_g$, $t_{visc}\simeq$ 60 days. It becomes $\sim$ 2000 days for $R=10^5 R_g$.}. 
 This is much longer than the few days duration for a typical transition between a hard (windless) state to a soft (windy) state (e.g. \citealt{dun10}).  To explain the observations, the communication between the X-ray corona and the wind must therefore be much faster than the accretion time scale. It has been pointed out that over-ionization, as a consequence of the much harder illuminating  spectral  energy  distribution during hard states might play a role (see e.g. \citealt{shi19}) but in several instances it may not be sufficient (e.g. \citealt{nei12,pon15})\\

It has been recently understood that the spectral state of the central source has a huge impact on the thermal stability of the wind and its ionization state \citep{cha13,cha16,bia17,hig20}. When illuminated by a soft state spectrum, the wind is always thermally stable regardless of its ionization parameter. When the spectrum is hard, however, large parts of the stability curve are thermally unstable, in particular in the ionization domain required for the production of Fe XXV and Fe\ XXVI ions. The timescale for the thermal instability to take place depends on the cooling timescale of the gas and, for the expected wind density (see Sect. \ref{sectSED}), is quite fast ($<$ day, see e.g. {{\citealt{gon07,bia17}}}). 
{{These arguments}} support the importance of {{the thermal instability}} process in the changes of the wind ionization state {(but see Sect. \ref{caveat} for potential caveats)}. In this context, the disk wind could be present at all times (as recently suggested by the observation of near-IR emission lines all along the outburst of the black hole transient MAXI J1820+070, \citealt{san20}), but will be detectable only in the soft state, when the relevant (for Fe\ XXV and Fe\ XXVI) range of ionization parameter corresponds to stable thermal equilibrium.\\

{{Stability curves are very useful to study more precisely the role of the thermal stability on the wind properties. They represent the photoionization equilibrium states of a plasma illuminated by a broad band Spectral Energy Distribution (SED).}} The computation of these stability curves is however very sensitive to the spectral shape of the illuminating SED (e.g. \citealt{cha13,bia17}). The precise study of the wind thermal stability during an outburst, and especially during the spectral transitions, then requires the knowledge of the spectrum in the broadest energy range possible.  However, Galactic hydrogen absorbs the soft X-rays ($<$ 0.1 keV) and instrumental limitations prevent access to a precise spectra for hard X-rays ($>$ 50 keV). With the typical 0.5-10 keV energy range (from e.g. {\it Chandra} or {\it XMM}), extrapolating from phenomenological models is hazardous and could lead to unphysical broad band spectral shapes and, consequently, erroneous stability curves. With the lack of daily broad band monitoring of XrB, the best alternative is the use of physically motivated models to obtain reasonable extrapolations where the data are missing.\\

This is the purpose of the present paper. We discuss the expected evolution of the disk wind properties of an XrB by using the detailed observations of the most famous (and thus best monitored) one i.e. GX 339-4. No wind signature has been detected in this object potentially due to the low inclination of our line-of-sight (\citealt{pon12}, but see \citealt{mil15} for the marginal detection of He-like Fe\ XXV and H-like Fe\ XXVI emission lines during a high/soft state of this source, potential signatures of a wind). {{Even if GX 339-4 is not the prototype of a windy XrB, we believe that its spectral behavior along an outburst can be used, as representative, to trace the one followed by any XrB. In consequence, its SED can  be used to trace the expected evolution of the stability curve of a windy source.}}
GX 339-4 entered several times in outburst in the last 20 years and for which a large amount of data exist (see e.g. the WATCHDOG all-sky database \citealt{tet16}). Its 2010--2011 outburst for example was very well followed in X-rays and radio. It was observed by {\it RXTE} for more than 400 days with almost one observation every day. \cite{mar19} recently reproduced the spectral shapes of each observation during the entire outburst with the Jet Emitting Disk (JED) model \citep{fer06a,mar18a,mar18b}. {{The corresponding best-fit SEDs are well constrained by the RXTE/PCA observations.  While model dependent, they are however physically motivated and give consistent extrapolations outside the energy window of RXTE}}.
{{We thus make the choice to}} use these SEDs to compute the wind stability curves and follow their evolution day by day, {{keeping in mind however their model-dependency}}. Assuming GX 339-4 is a typical XrB, this evolution of the stability curves is expected to be quite generic for all XrBs. Moreover, the very good sampling of the hard-to-soft transition of the 2010-2011 outburst of GX 339-4 is optimal to illustrate our purposes. { None of the sources which show absorption lines in its spectra have as extensive coverage of SEDs as a function of state evolution}. \\

This paper is organized as follows. {{In Sec. \ref{sectSED} we explain how we compute the stability curves}}. Then, we present and discuss in details their evolution all along the outburst in Sec. \ref{stabcurvevol}. Our results suggest significant changes in the disk wind properties during the state transitions, discussed in Sec. \ref{windprop}.  We propose that the disk illumination plays a major role in the wind observable properties. We conclude in Sect. \ref{discussion}.

%
%
%
%

\section{{{Stability curves computation}}}
\label{sectSED}
\subsection{{{Methodology}}}
{{Stability curves are generally plotted in a $\log T-\log \xi/T$ diagram (e.g. \citealt{kro81}), where $T$ is the gas temperature and $\xi$ its ionization parameter\footnote{We use here the original \cite{tar69} definition of the ionisation parameter $\xi=L/nr^2$, where $L$ is the total luminosity in ionizing photons, $n$ is the hydrogen density and $r$ is the distance of the gas from the illuminating source.}. In this diagram, an equilibrium state where the slope of the curve is positive is thermally stable, while it is unstable if the slope is negative, and the gas is then expected to rapidly collapse into a different stable equilibrium state.}}

{{As said in the introduction, we used the best fit SED obtained by \cite{mar19} in the Jet-Emitting Disk model framework to compute the stability curves. These SED reproduce the {\it RXTE/PCA} data of GX 339-4 observed during its 2010-2011 outburst. The JED framework and the way the SED are computed are described in a series of papers \citep{fer06a,mar18a,mar18b} to which the reader is referred for more details. We have also reported in Appendix \ref{app1} a few informations in this respect. To summarize, the JED framework assumes the presence of a Jet Emiting Disk in the inner region of the accretion flow, for disk radii $< R_{tr}$, and a Standard Accretion Disk (SAD) beyond $R_{tr}$. The total SED is mainly characterized by $R_{tr}$ and the accretion rate $\dot{M}$ reaching the inner compact object. The high energy emission (above 1-2 keV) is dominated by the JED emission while below $\sim$1 keV it is dominated by the emission of the standard accretion disk. Both JED and SAD emissions depend on $R_{tr}$ and  $\dot{M}$.}} {The JED emission is generally dominated by the Compton process, the seed photon field being a mix of local bremsstrahlung and synchrotron emission and non-local emission from the SAD (see \cite{mar18b}. A few SED examples are detailed in Fig. \ref{SEDcompton}).} \\

\begin{figure}[t]
\includegraphics[width=\columnwidth]{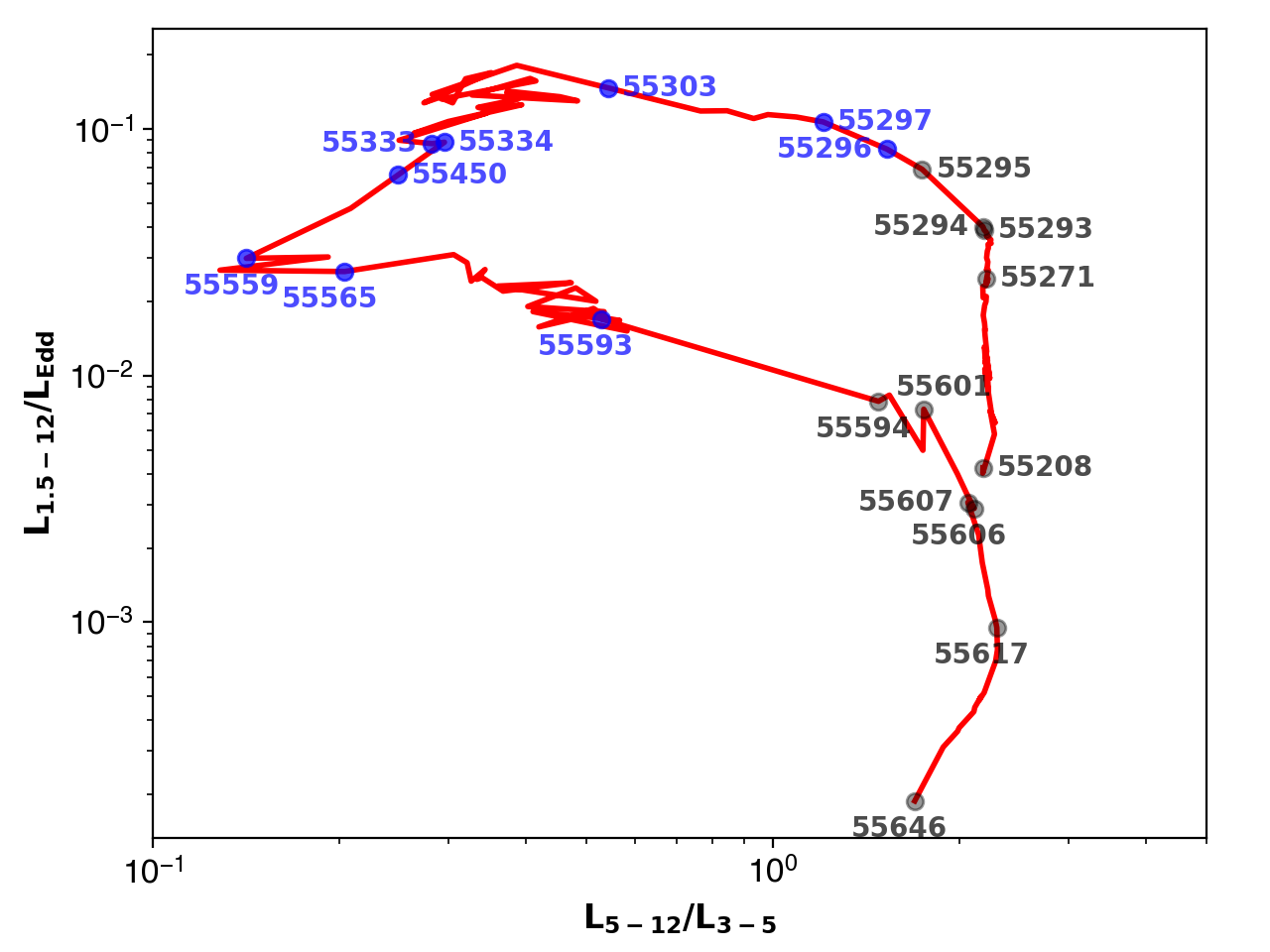}
 \caption{HID of the 2010-2011 outburst of GX 339-4. The position of the observations reported in Tab. \ref{param} are indicated with blue and black colors and are labelled on the HID with their MJD. The black (resp. blue) MJD correspond to observations whose range of ionisation parameter consistent with Fe\ XXV and Fe\ XXVI ions is in an unstable (resp. stable) part of the photoionisation stability curves. We have distributed the MJD observations in terms of modeled SED and associated stability curves in Fig. \ref{figStabCurvTrans}.}
\label{figStabCurv}
\end{figure}

{{\cite{mar19} show that $R_{tr}$ and  $\dot{M}$, and consequently the broad band JED-SAD SEDs, are quite well constrained all along the outburst thanks to the RXTE/PCA data.}} We have reported in Tab. \ref{param} the parameters $R_{tr}$ and $\dot{M}$ as well as the corresponding bolometric and ionizing (i.e. between 1 and 1000 Rydbergs) luminosities (in Eddington units) of a few {{best fit}} SEDs at different MJD along the 2010-2011 outburst of GX 339-4. 
We have reported in Fig. \ref{figStabCurv} the  Hardness-Intensity Diagram (HID) of this outburst  with the position of the different observations reported in Tab. \ref{param}. The corresponding SEDs are plotted in the left part of Fig. \ref{figStabCurvTrans} and will be discussed later. 

These SEDs are then used in \textsc{cloudy} 17.01 (last described in \citealt{ferl13}) to compute the corresponding stability curves all along the outburst. \begin{table}
\begin{center}
\begin{tabular}{cccccc}
\hline
MJD  & $R_{tr}$  & $\dot{M}$ & $L_{1-1000 Ry}$ & $L_{1-100 keV}$ & $L_{bol}$\\
  & ($R_{g}$)  & ($\dot{M}_{Edd}$) & ($L_{Edd}$) & ($L_{Edd}$) & ($L_{Edd}$)  \\
\hline
55208 & 27.1 & 0.63 & 0.019 & 0.014 & 0.043\\ 
55271 & 23.1 & 1.57 & 0.062 & 0.083 & 0.174\\ 
55293 & 23.1 & 2.42 & 0.094 & 0.130 & 0.248\\
55294 & 21.6 & 2.42 & 0.099 & 0.132 & 0.249\\ 
55295 & 9.6 & 2.34 & 0.177 & 0.157 & 0.264\\ 
55296 & 8.4 & 2.51 & 0.210 & 0.170 & 0.286\\ 
55297 & 7.2 & 2.80  & 0.261 & 0.194 & 0.324\\ 
55303 & 5.0 & 2.80  & 0.330 & 0.217 & 0.352\\ 
55333 & 2.2 & 0.75  & 0.168 & 0.114 & 0.168\\ 
55334 & 2.0 & 0.70  & 0.166 & 0.115 & 0.166\\ 
55450 & 2.0 & 0.55  & 0.129 & 0.087 & 0.129\\ 
55559 & 2.2 & 0.33  & 0.073 & 0.043 & 0.073\\
55565 & 2.5 & 0.34  & 0.070 & 0.040 & 0.070\\
55593 & 3.0 & 0.27  & 0.052 & 0.029 & 0.055\\
55594 & 4.9 & 0.27  & 0.034 & 0.018 & 0.041\\
55601 & 6.6 & 3.33  & 0.033 & 0.019 & 0.044\\
55606 & 15.3 & 0.35 & 0.017 & 0.009 & 0.028 \\
55607 & 15.6 & 0.37 & 0.017 & 0.009 & 0.029 \\
55617 & 24.2 & 0.15 & 0.005 & 0.003 & 0.008\\
55646 & 40.1 & 0.03 & 0.0006 & 0.0004 & 0.001\\
\hline
\end{tabular}
\caption{MJD, transition radius and accretion rate of  {{some of the best fit SED obtained by \cite{mar19} to reproduce the {\it RXTE/PCA} observations of the 2010-2011outburst of GX 339-4. We have also reported the corresponding luminosities between 1 and 1000 rydbergs and between 1 and 100 keV as well as the bolometric luminosity}}. The Eddington luminosity $L_{Edd}=8.7 \times 10^{38}$ erg.s$^{-1}$ assuming a black hole mass of 5.8 $M_{\odot}$. The distance of GX 339-4 is taken equal to 8 kpc. \label{param}}
\end{center}
\end{table}
We adopted Solar abundances (as in Table 7.1 of \textsc{cloudy} documentation\footnote{The \textsc{cloudy} document can be found at this \href{http://web.physics.ucsb.edu/~phys233/w2014/hazy1_c13.pdf}{link}}), a typical disk wind density $n=10^{12}$ cm$^{-3}$ (see next Sect. \ref{caveat} for a discussion on the effect of the density), a turbulence velocity $v_{turb}=500$ km s$^{-1}$ and a column density $\log(N_\mathrm{H}/\mathrm{cm}^{-2})=23.5$,  as in \cite{bia17}. 

\subsection{{{Caveats}}}
\label{caveat}
{{While the SED have strong impact on the stability curves, other parameters can also affect the heating/cooling equilibrium and modify the stability curve shape. For instance, \cite{bia17} looked at the effect of the chemical abundances and they are marginal. \cite{cha13} and \cite{bia17} looked at the effect of the plasma density and only high densities (above 10$^{16}$ cm$^{-3}$) have significant impact on the stability curve shape. In the case of hard state SED  the slope of the stability curves can even change and the branch consistent with Fe\ XXV and Fe\ XXVI ions can move from thermally unstable to thermally stable conditions (see Fig. 5 of \citealt{bia17}). Such high densities are however not consistent with the (admittedly rare) observational constraints (e.g. \citealt{kal09,sch08}) nor expected from disk wind models in XrB (e.g. \citealt{cha16}, \citealt{hig17}). Our choice of a wind density of $n=10^{12}$ cm$^{-3}$ in our simulations is then coherent with the fact that stability curves do not change drastically for a wind density varying between $10^8-10^{16}$ cm$^{-3}$  (see \citealt{cha13,bia17}, and see also Appendix \ref{effectdens}).}}

{{The Cloudy code also assumes an isolated and static plasma which is obviously not the case for a large scale wind. The effects like thermal conduction or adiabatic coolings are thus not taken into account in our computation. Numerical simulations of thermal disk winds indicate however that thermal conduction should have a small effect on the heating/cooling equilibrium (e.g. \cite{hig17} for thermal winds). Concerning adiabatic coolings, they can also be neglected as long as the time to reach photoionisation equilibrium is, locally, shorter than the dynamical timescale of the wind. While the observed line widths generally imply that the gas is supersonic, such conditions are actually verified in MHD disk wind solutions like those discussed by \cite{cha16}. Actually, even if the plasma wind velocity could be supersonic, its dynamical timescale (see an estimate given by \cite{gon07} in their Eq. 8), is  expected to be significantly longer than the other timescales involved in the radiative equilibrium computation (e.g. ionisation time, recombination time, thermal time, see \cite{gon07}).

While these arguments suggest that the shape of the stability curves discussed in this paper should not strongly depend on our physical and numerical assumptions, a precise estimates would require a detailed modeling of the disk wind physical properties and dynamics {like in the recent work by \cite{dan20,wat21} in the case of thermal disk winds. While this is an important step that should be done also in the case of magnetically driven disk winds,} this is however out of the scope of the present paper. Some caution is nevertheless required and our results must then be understood with these limitations in mind.}} 

\section{Stability curve evolution}
\label{stabcurvevol}
The  stability curves corresponding to the MJD reported in Tab. \ref{param} are plotted in the right panels of Fig. \ref{figStabCurvTrans} in the $\log T$-$\log \xi/T$ plane. In this plane, the parts of the stability curves with positive (resp. negative) slopes are thermally stable (resp. unstable). The highlighted areas on each curve correspond to the range of ionisation parameter consistent with Fe\ XXV and Fe\ XXVI ions\footnote{More precisely, they correspond to a width in $\log\xi$ at 90 per cent of the peak of their ionic fractions $f_{ion}$, $f_{ion}(\xi)$ depending on the SED. We have reported a few examples of ion fraction in the Appendix Fig. \ref{ionfraction}.
} and which can go from $\sim10^3$ to $\sim10^6$ erg s$^{-1}$ cm depending on the SED. They are colored in gray (resp. blue) if they are located on an unstable (resp. stable) part of the stability curve. Clearly, the gray colored ones correspond to the observations on the right part of the HID, i.e. from MJD 55208 to 55295 and from MJD 55594 to 55646, which are either hard or hard-intermediate states of the outburst. On the contrary, all the other observations (in soft and soft-intermediate states) have highlighted areas on stable branches. This is in agreement with recent studies (e.g. \citealt{cha13,cha16,bia17}) and, as discussed in the introduction, this would explain the non observations of Fe\ XXV-Fe\ XXVI absorption lines in hard states while a wind could be still present.

\begin{figure*}
\begin{center}
\begin{tabular}{cc}
\setlength{\unitlength}{1cm}
\begin{picture}(7.5,5.5)
\put(0,0.1){\includegraphics[height=5.5cm,width=7.7cm]{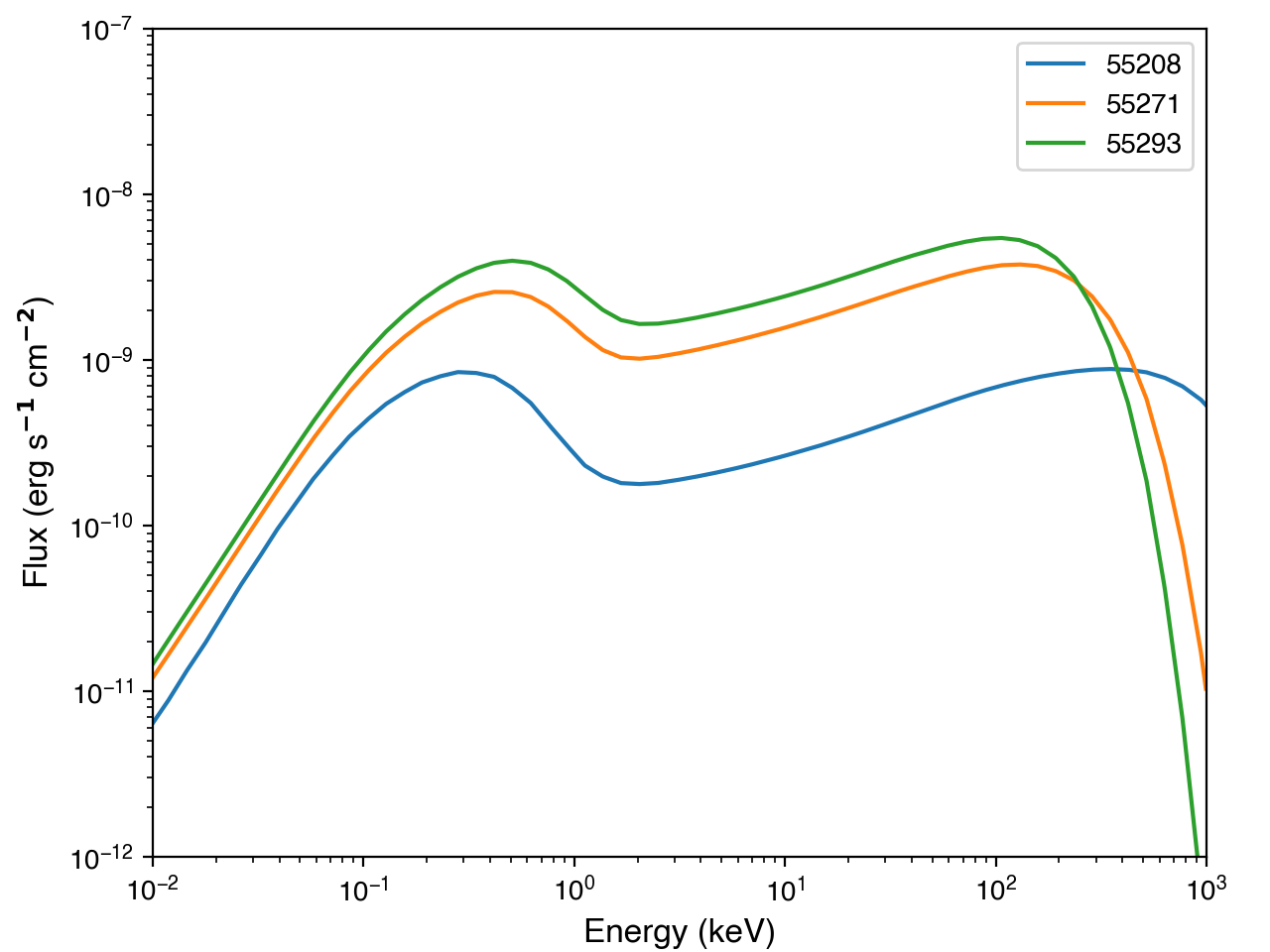}}
\put(1.1,5.1){\large a)}
\end{picture}
 &
\setlength{\unitlength}{1cm}
\begin{picture}(7.5,5.5)
\put(0,0.1){\includegraphics[height=5.5cm,width=7.7cm]{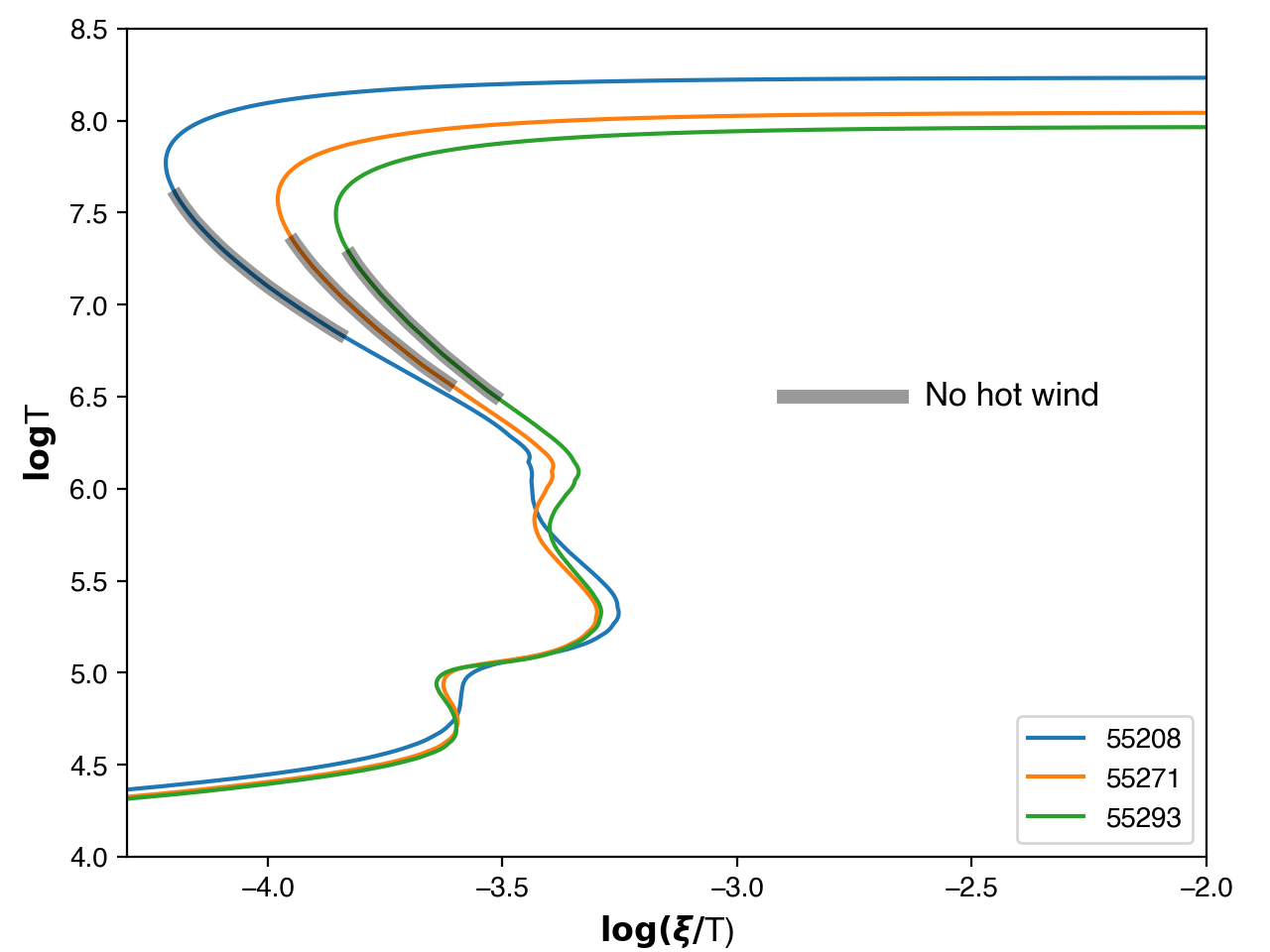}}
\put(1.1,5.1){\large b)}
\end{picture}\\

\setlength{\unitlength}{1cm}
\begin{picture}(7.5,5.5)
\put(0,0){\includegraphics[height=5.5cm,width=7.7cm]{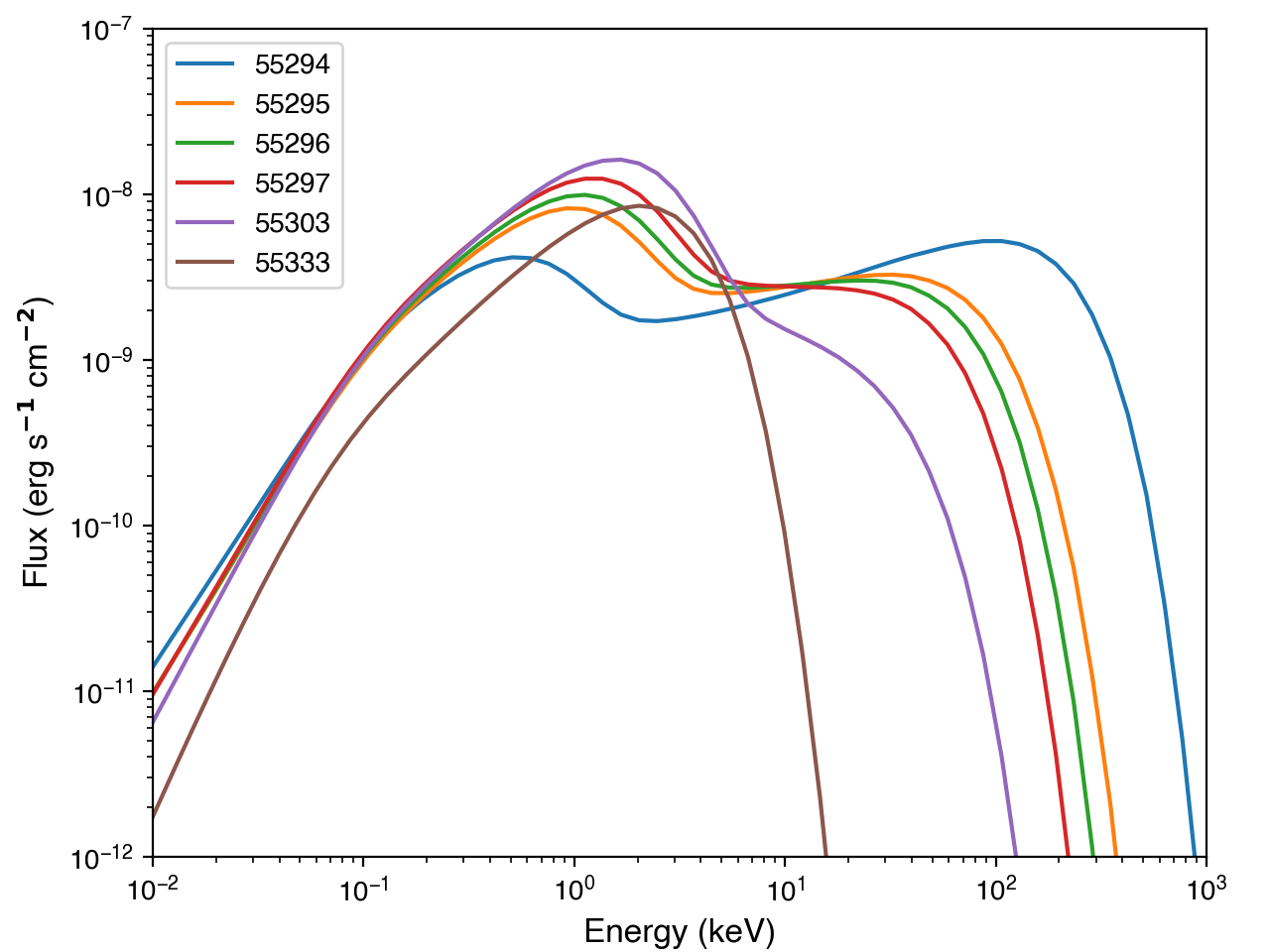}}
\put(6.5,5.){\large c)}
\end{picture}
 &
\setlength{\unitlength}{1cm}
 \begin{picture}(7.5,5.5)
\put(0,0){\includegraphics[height=5.5cm,width=7.7cm]{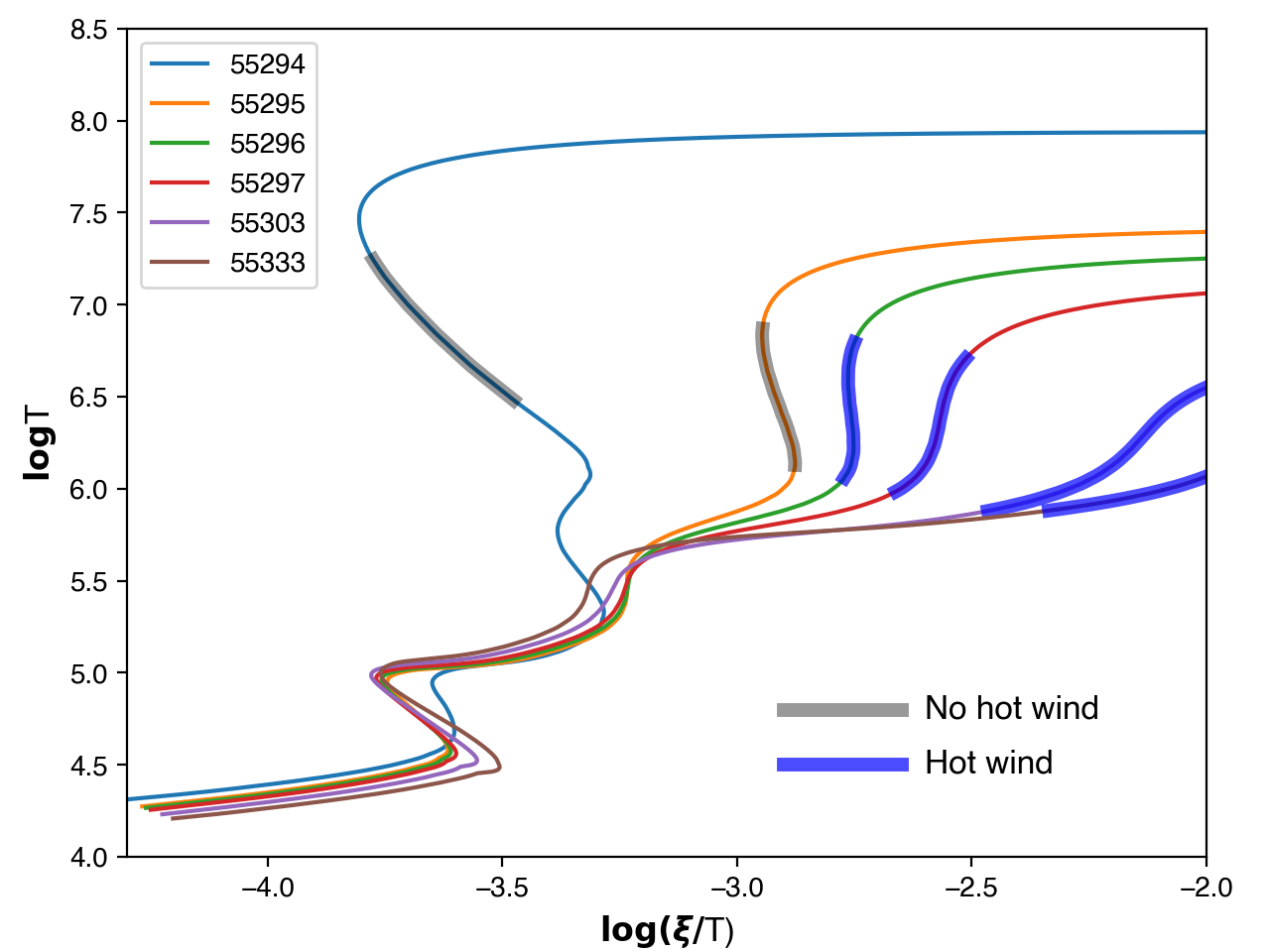}}
\put(6.5,5.){\large d)}
\end{picture}\\

\setlength{\unitlength}{1cm}
\begin{picture}(7.5,5.5)
\put(0,-0.1){\includegraphics[height=5.5cm,width=7.7cm]{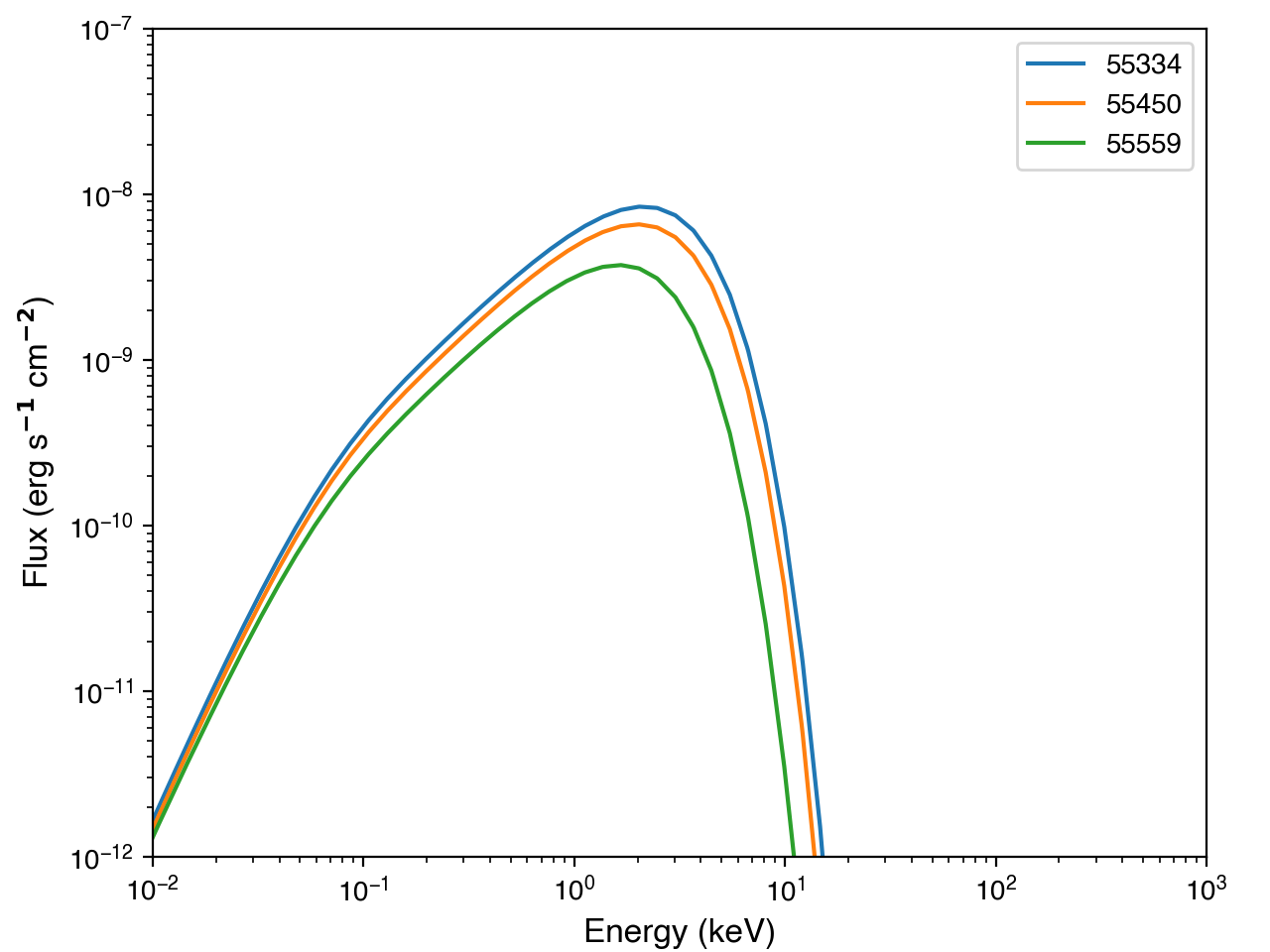}}
\put(1.1,4.8){\large e)}
\end{picture}
 &
\setlength{\unitlength}{1cm}
 \begin{picture}(7.5,5.5)
\put(0,-0.1){\includegraphics[height=5.5cm,width=7.7cm]{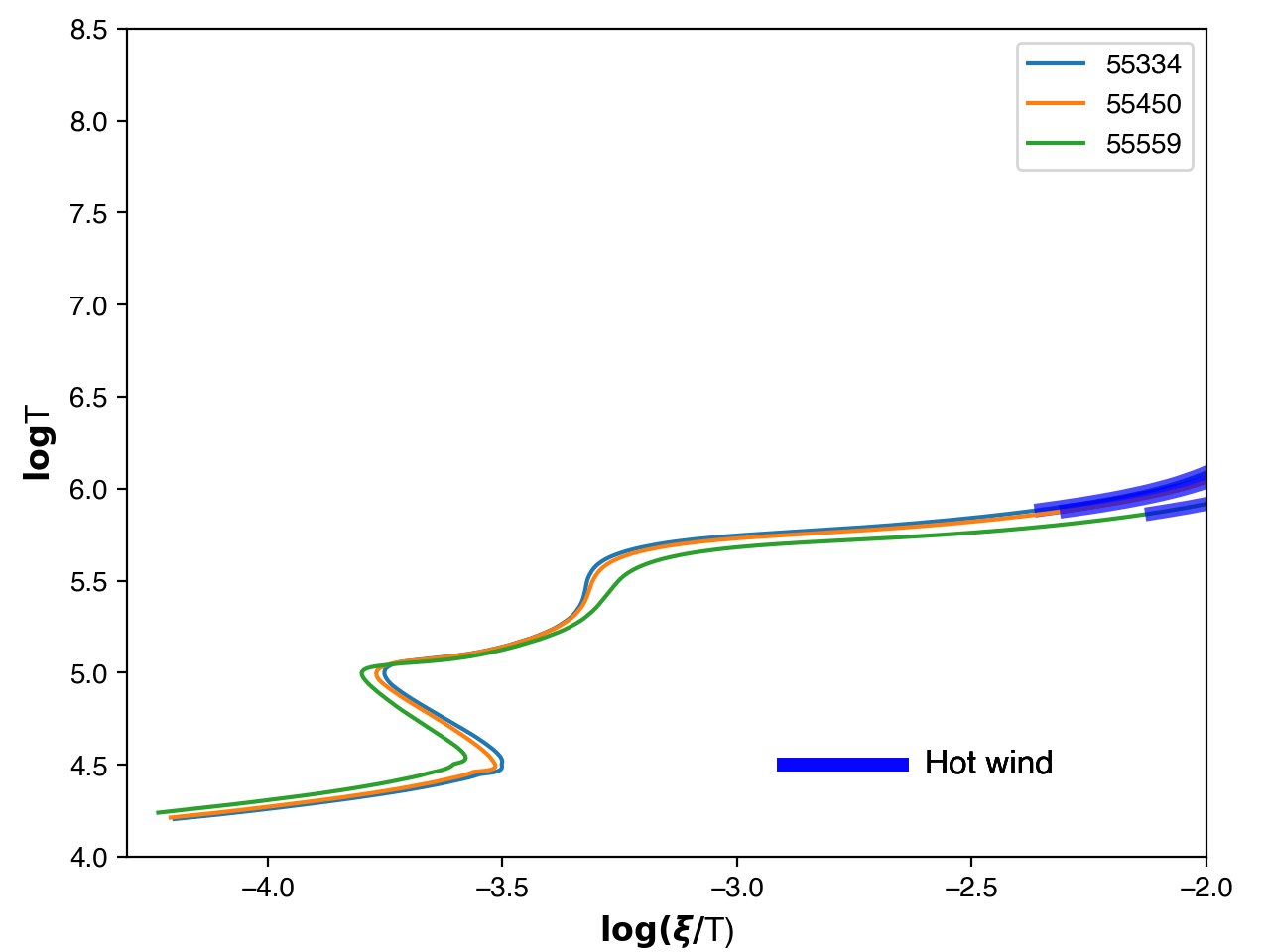}}
\put(1.1,4.8){\large f)}
\end{picture}\\

\setlength{\unitlength}{1cm}
\begin{picture}(7.5,5.5)
\put(0,-0.2){\includegraphics[height=5.5cm,width=7.7cm]{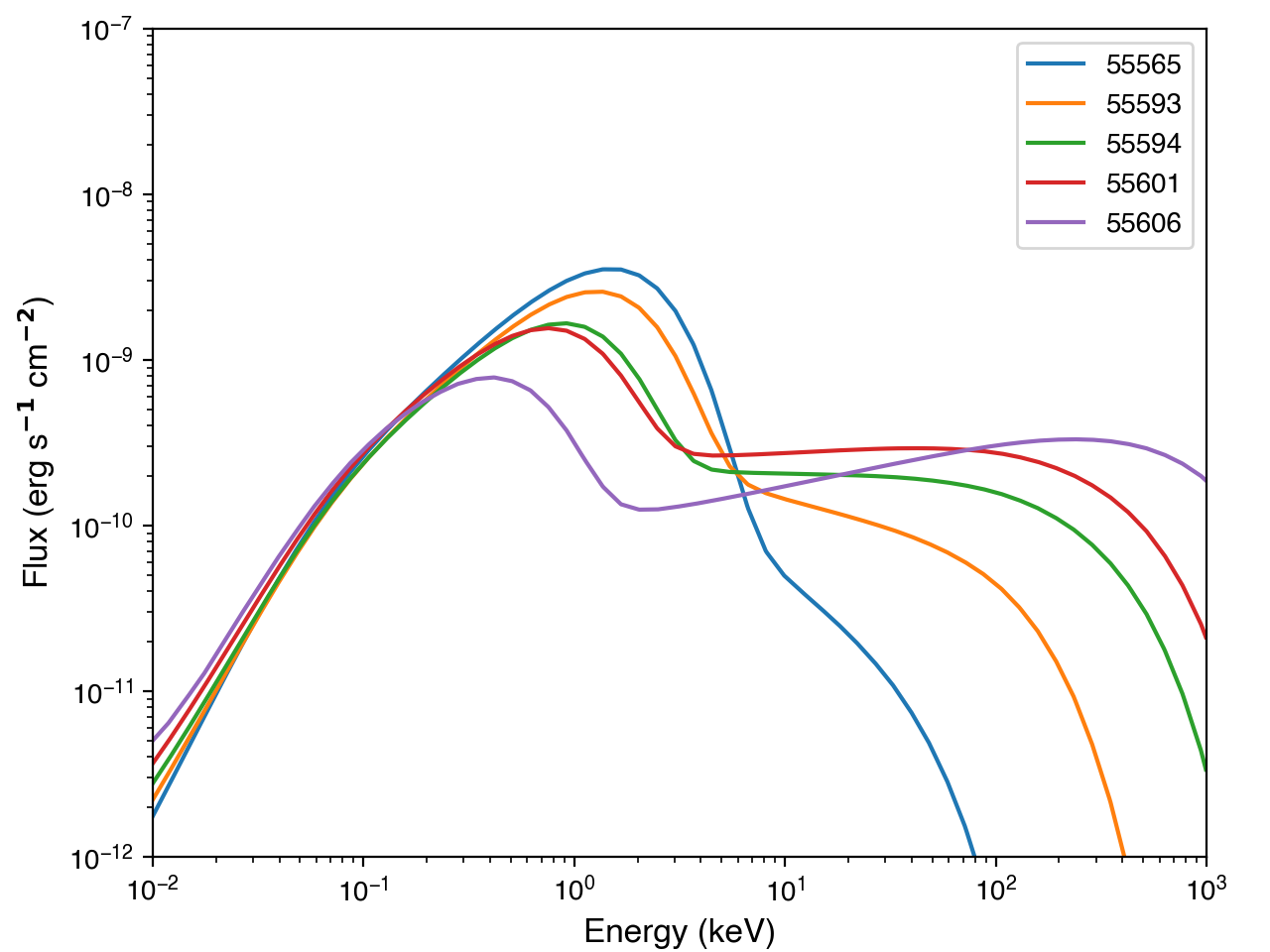}}
\put(1.1,4.8){\large g)}
\end{picture}
 &
\setlength{\unitlength}{1cm}
 \begin{picture}(7.5,5.5)
\put(0,-0.2){\includegraphics[height=5.5cm,width=7.7cm]{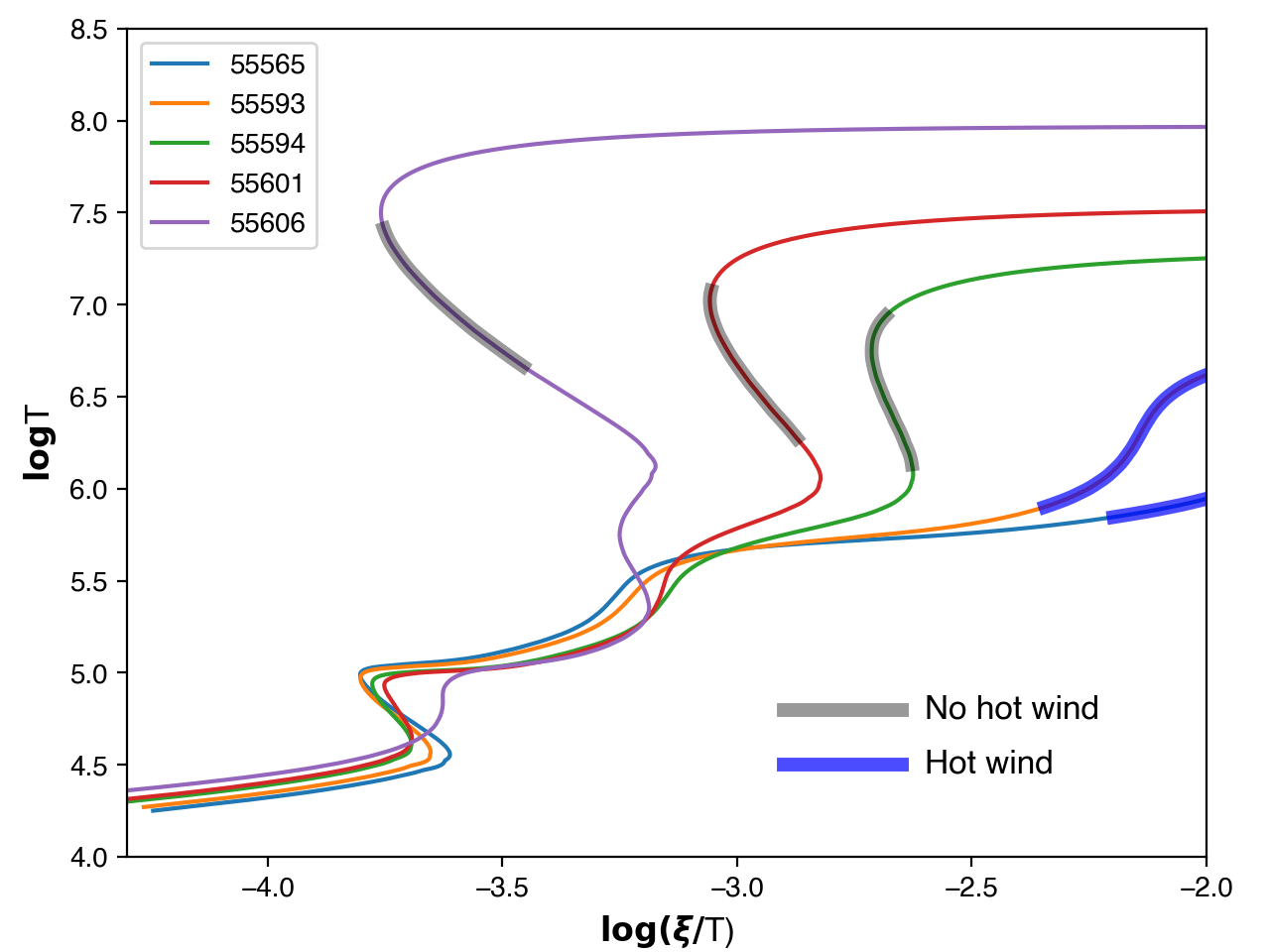}}
\put(6.5,4.8){\large h)}
\end{picture}\\
\end{tabular}
 \caption{SED (left) and corresponding photoionisation stability curves (right) between MJD 55208 and 55293 (hard state, a) and b)), MJD 55294 and 55303 (hard-to-soft transition, c) and d)), MJD 55304 and 55592 (soft state, e) and f)) and  MJD 55593 and 55606 (soft-to-hard transition, g) and h)). The stability curves are labeled with their corresponding MJD. The highlighted (blue and gray) areas on each curve correspond to the range of ionisation parameter consistent with Fe\ XXV and Fe\ XXVI. They are colored in blue (resp. gray) if they are located on a stable (resp. unstable) part of the stability curve.
}

\label{figStabCurvTrans}
\end{center}
\end{figure*}
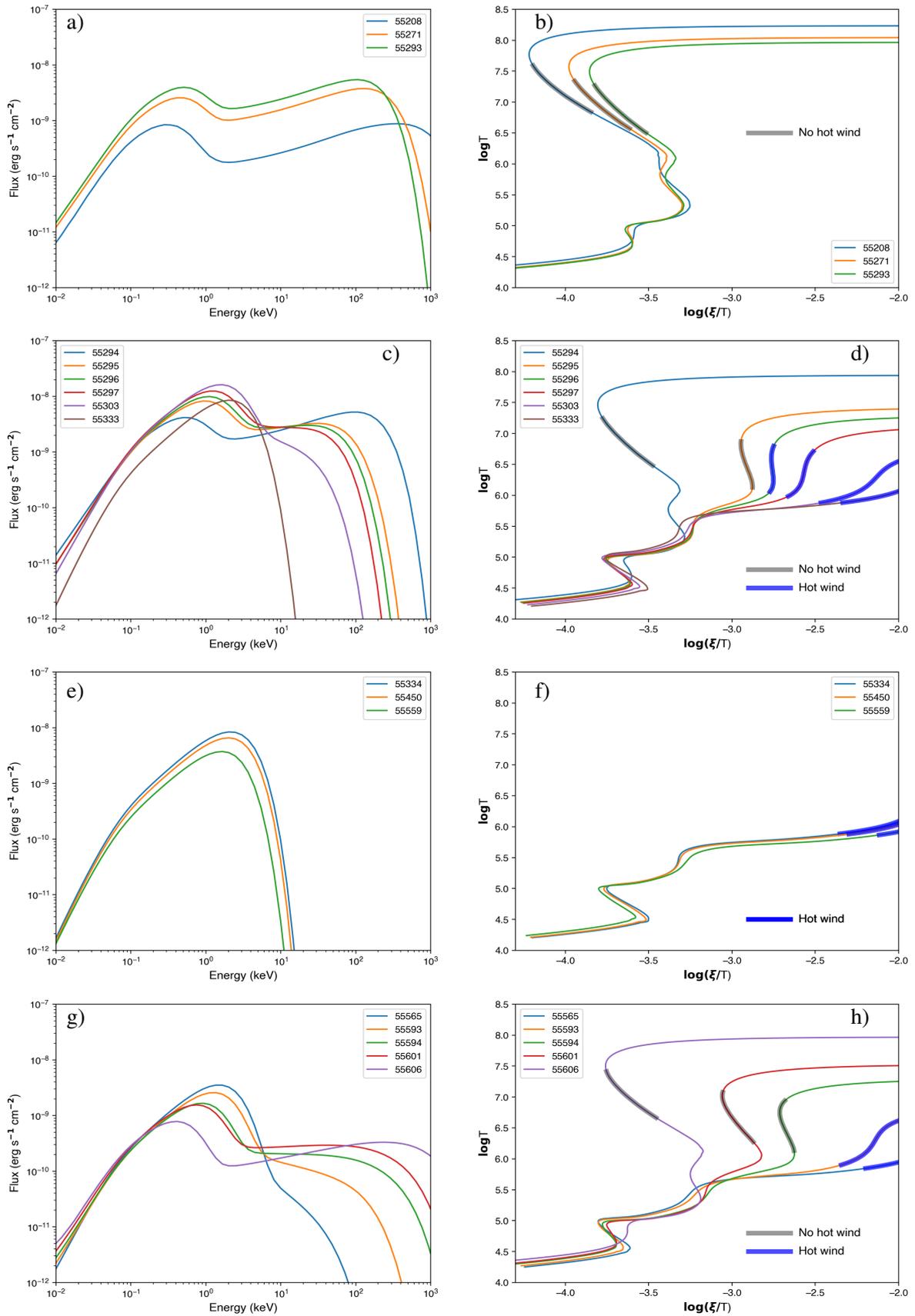

\setcounter{figure}{1}
\begin{figure*}[t]
\begin{center}
\begin{tabular}{cc}
\setlength{\unitlength}{1cm}
\begin{picture}(7.5,6)
\put(0,0){\includegraphics[height=5.5cm,width=7.7cm]{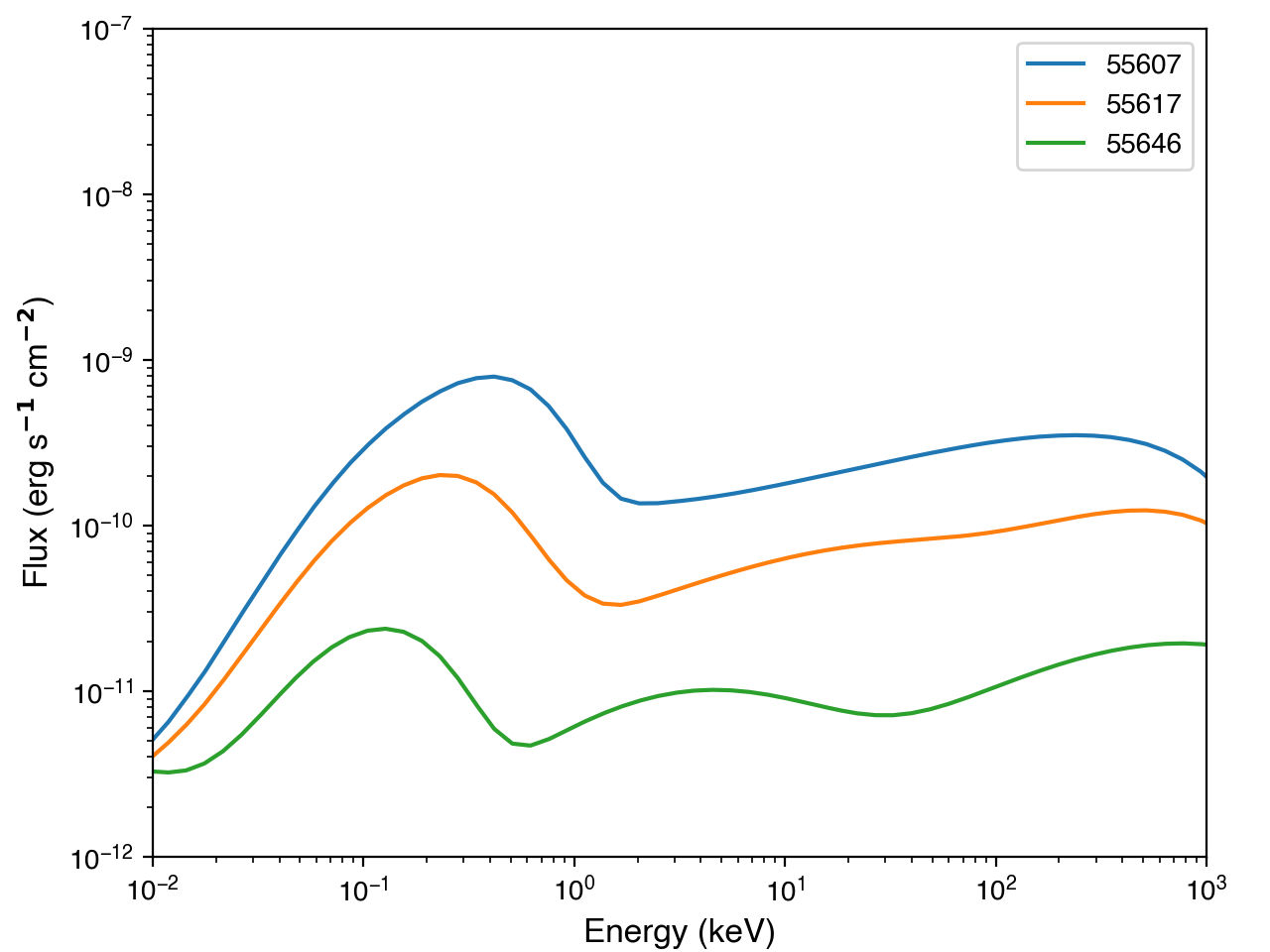}}
\put(1.1,4.9){\large i)}
\end{picture}
 &
\setlength{\unitlength}{1cm}
 \begin{picture}(7.5,6)
\put(0,0){\includegraphics[height=5.5cm,width=7.7cm]{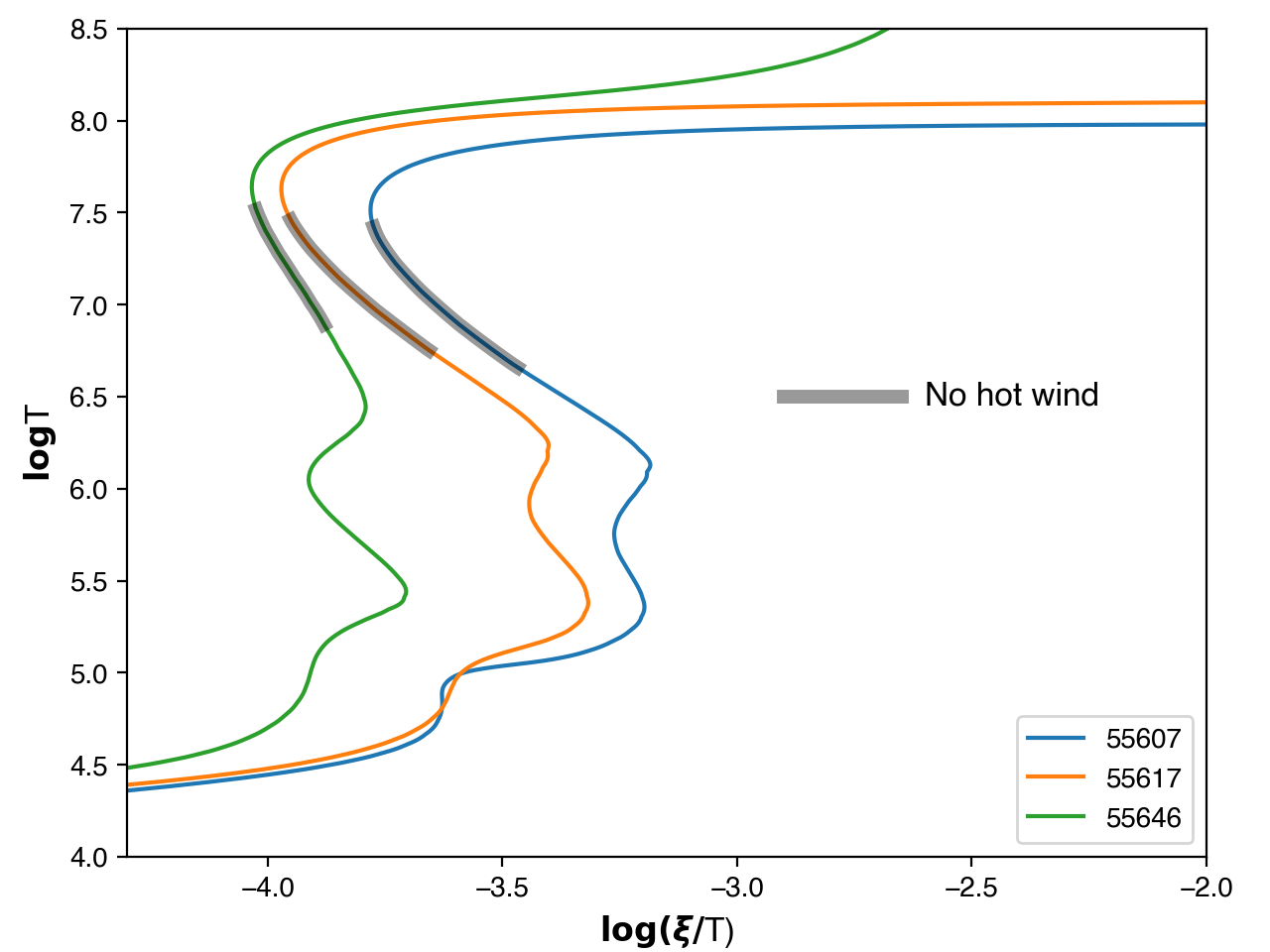}}
\put(1.1,4.9){\large j)}
\end{picture}\\
\end{tabular}
 \caption{{\bf cont.} i) Photoionisation stability curves and j) corresponding SED between MJD 55607 and 55646 (hard-to-quiescent state).}
\label{figStabCurvTransb}
\end{center}
\end{figure*}
\label{tempevol}
The daily observations of the 2010-2011 outburst of GX 339-4 and the best fit SEDs obtained with the JED model allow us to follow the evolution of the stability curves all along the outburst. We describe below this evolution during the different spectral stages of the source. While not unique, we follow the spectral state definitions of \cite{mar19}.
\subsection{Hard state: from MJD 55208 to MJD 55293} 
\label{HS}
The first {\it RXTE} observation starts at MJD 55208 when the source was already in a quite bright ($L\simeq 2\%L_{Edd}$) hard state. GX 339-4 stayed in the hard state until around MJD 55293 when it started transitioning to the soft state. The 1-1000 Ry luminosity increases by a factor 5-6 during this period (see Tab.~\ref{param}). The three best-fit SEDs obtained with the JED model for MJD 55208, 55271 and 55293 are reported in Fig. \ref{figStabCurvTrans}a and the corresponding stability curves in Fig. \ref{figStabCurvTrans}b. The SEDs are typical of hard states, dominated by a hard X-ray cut-off power law above $\sim$1-2 keV (produced by the inner JED) and a weak disk black-body component below, produced by the outer SAD.  The power-law slope appears roughly constant around 1.4-1.5 and the cut-off energy decreases when the luminosity increases. The disk black body emission increases in flux and temperature in agreement with the decrease of the transition radius $R_{tr}$ between the JED and the SAD while the accretion rate increases (see Tab. \ref{param}).

The computed stability curves are quite similar for the three MJD 55208, 55271 and 55293. In the three cases, the temperature range that agrees with the right ionization range for Fe\ XXV and Fe\ XXVI is unstable. Actually, if the wind exists during this period, its temperature has to be in one of the ranges 
where the stability curves have a positive slope i.e. either at very high temperature ($>3\times10^{7}$ K) or in localized regions around $10^6$ K, $2\times10^5$ K or at lower temperature $<3\times10^{4}$ K. In these temperature ranges, the wind would not produce any Fe\ XXV and Fe\ XXVI absorption lines (see Sec. \ref{lowtemp}).

\subsection{Hard-to-soft transition: from MJD 55294 to MJD 55333}
From MJD 55294 the source starts the transition to the soft state, i.e. the path in the HID turns to the left. 
This transition lasts about 40 days, between MJD 55294 to around MJD 55333. During this period the SEDs still show a power law component in the X-rays but the slope increases from $\sim$1.5 to 2.2 and the high energy cut-off decreases all along the transition (see Fig. \ref{figStabCurvTrans}c). On the other hand, the disk component increases in flux and temperature. At MJD 55333, at the end of the transition, the SED is only dominated by the disk component. 

We have reported the corresponding stability curves in Fig. \ref{figStabCurvTrans}d. 
During this hard-to-soft transition, the stability curves change drastically and at the end of the transition (after MJD 55296), the stability curves become thermally stable in almost all the temperature range used in the CLOUDY computations\footnote{except in a narrow interval around $3\times10^{4}-10^5$ K, a temperature range which is irrelevant (because too low) for Fe\ XXV and Fe\ XXVI}. This includes the ionization range compatible with Fe\ XXV and Fe\ XXVI ions. 

This change occurs in a couple of days (between MJD 55295 and MJD 55297) while the changes of the SED are quite minor. Within a day, from MJD 55295 to MJD 55296, the wind  becomes able to produce Fe\ XXV and Fe\ XXVI absorption lines. 
While the bolometric luminosity varies by only a factor 1.4 during the transition, with a maximum around MJD 55303, the ionizing luminosity varies by a significantly larger factor $\sim$3.3 due to the increase of the disk flux which dominates the 1-1000 Ry energy range.

%

\subsection{Soft state: from MJD 55334 to MJD 55559.} 
\begin{figure*}[t]
\begin{center}
\begin{tabular}{cc}
\includegraphics[width=0.5\textwidth]{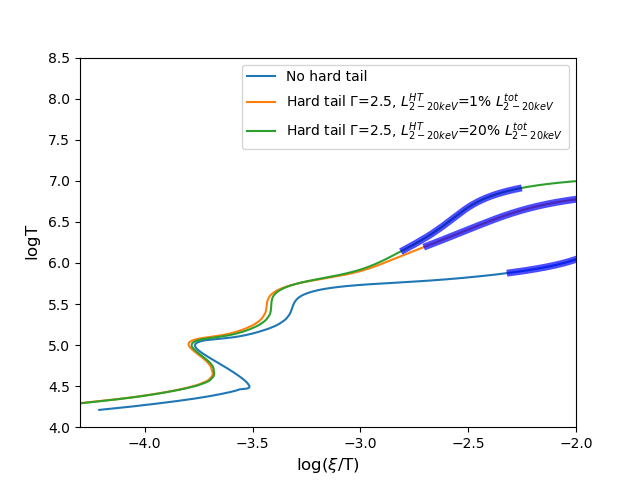} &
\includegraphics[width=0.5\textwidth]{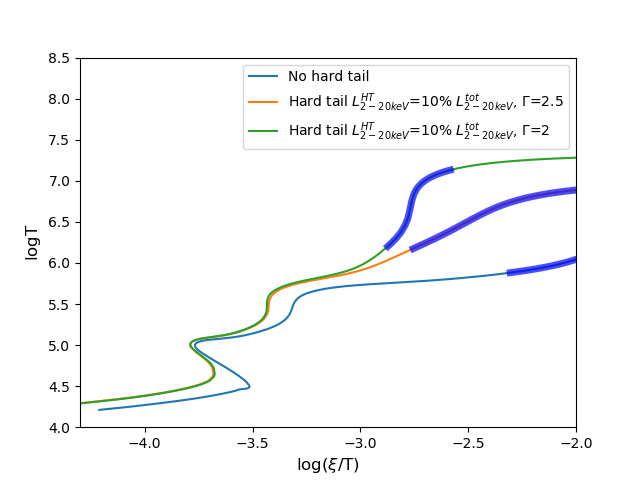} 
\end{tabular}
\end{center}
\caption{Effect of the hard tail on the photoionisation stability curves. Application to MJD 55410. {\bf Left:}  hard tails with constant photon indexes $\Gamma$=2.5 but with different fluxes contributing to 1 and 20\% of the 2-20 keV flux. {\bf Right:} hard tails with different photon index ($\Gamma=$2 and 2.5) but a constant flux contributing to 10\% of the 2-20 keV flux.}
\label{fighardtail}
\end{figure*}

In the soft state, which lasts here about 225 days, the SED are mainly characterized by a disk component in this state (see Fig. \ref{figStabCurvTrans}e). The stability curves are very similar to one another during this entire state (see Fig. \ref{figStabCurvTrans}f) while the 1-1000 Ry luminosity decreases by a factor 4-5. The ionization range compatible with Fe\ XXV and Fe\ XXVI ions is always in a stable part of the stability curves, meaning that the associated absorption lines can be present. The presence of a hard tail can however significantly impact the thermal equilibrium of the wind (see Sect. \ref{hardtail} and Fig. \ref{fighardtail}).

\subsection{Soft-to-hard: from MJD 55561 to MJD 55606.}
\label{softtohard}
At MJD 55559, the path of GX 339-4 starts to turn right in the HID, the source transiting back to the hard state. This transition lasts about 40 days until around MJD 55606. We have reported a few best fit SEDs during this transition in Fig. \ref{figStabCurvTrans}g. The disk component decreases in flux and temperature while the X-ray power law hardens and the high-energy cut-off increases. The  1-1000 Ry and bolometric luminosities decrease by a factor $\sim$2.2 and $\sim$2.6 respectively during this transition.

We reported the corresponding stability curves in  Fig. \ref{figStabCurvTrans}h.
Like in the hard-to-soft transition, the shape of the stability curves changes drastically during the soft-to-hard transition. Moreover, and again similar to the hard-to-soft transition, the change from stable to unstable conditions for the ionization range compatible with Fe\ XXV and Fe\ XXVI ions occurs in a couple of days.  While at MJD 55592 the stability curve is similar to that of soft states, at MJD 55594 it becomes the same as hard states at the beginning of the outburst.  Moreover, the temperature ranges of thermal stability after the soft-to-hard transition are also similar to those before the hard-to-soft transition (see Sect. \ref{HS}). 

\subsection{Hard-to-quiescence: from MJD 55607 to MJD 55646. }
\label{quies}
After MJD 55607, GX 339-4 is back to the hard state until the end of the outburst. The bolometric luminosity decreases by more than a factor 30 during this period. The lower the luminosity the more bumpy the SED shape (see Fig. \ref{figStabCurvTransb}i). This is characteristic of the Compton process in a hot but tenuous plasma where each scattering order produces well separated spectral bumps. 

We reported the corresponding stability curves in Fig. \ref{figStabCurvTransb}j. They are typical of Hard State ones (see Sect. \ref{HS}) and no Fe\ XXV and Fe\ XXVI absorption lines are expected. The global shape of the stability curves seems to stiffen however when the source flux decreases, the ``S'' shape evolving to a ``step'' function where the upper hot and tenuous branch takes up the large part of the $\xi/T$ range. 

\subsection{Hard tail}
\label{hardtail}
The soft states can also present a low luminosity and steep power law at high energy (above $\sim$ 10 keV), the so-called hard tail, whose origin is currently unknown \citep{rem06}. It is generally badly constrained due to the lack of statistics, reaching up to a 50\% error on the photon index for example. \cite{mar19} includes this hard tail in the SED by adding a power-law component to the synthetic spectra each time the fitting procedure favors a pure blackbody emission (see \cite{mar18b} for a detailed discussion). The photon index of this power-law component is set to $\Gamma$ = 2.5 and it is normalized in order to contribute to a fixed fraction of the 2-20 keV energy range (typically between 1\% and 20\%, \citealt{rem06}). We report in the left of Fig. \ref{fighardtail} the stability curves for MJD 55410 assuming different hard tails with different flux (contributing to 1 and 20\% of the 2-20 keV flux) but with a fixed photon index ($\Gamma$=2.5). In the right panel, we assume hard tails with different spectral indices  ($\Gamma$=2.5 and 2) but a constant flux (contributing to 10\% of the 2-20 keV flux).  These different examples show that the presence of this hard tail can have some effects on the stability curves but is not sufficient to change a soft state-like into a hard state-like shape. Moreover its impact  on the thermal stability of the hot wind in the soft state should not drastically change the existence of a stable ionisation domain for Fe\ XXV and Fe\ XXVI. 

\subsection{{{Low temperature phases}}}
\label{lowtemp}
{All the stability curves plotted on Fig. \ref{figStabCurvTrans} and \ref{fighardtail} show low temperature domains (around $10^{4.5}$ and $10^5$ K) which are thermally stables. If part of the wind plasma is in this range of temperatures, absorption lines from low ionisation-level ions could also be observable. Of course this will strongly depend on the total column density of this low temperature plasma on the line of sight. For instance, in the case of the transient accreting neutron star AX J1745.6-2901, \cite{bia17} tested the case where the column density is equal to the one deduced from the Fe\ XXV and Fe\ XXVI absorption lines. This low temperature phase, if present on the line of sight, would then block all the radiation in the soft X-rays. On the other hand, \cite{bia17} estimate the filling factor of this phase to be very low. So this strong absorption would only be observed as a sporadic change of the persistent neutral absorption and, in the case of  AX J1745.6-2901, this was tentatively  associated to the dips observed in this source \cite{bia17}. We simulated with Cloudy cases with lower column densities where the absorption is less severe. No absorption lines are present above 1 keV, but some significant absorption is observed below 1 keV and strong absorption in the UV down to the IR. Note also that the ionisation parameter is quite low in these phases ($< 10^3$). This could correspond to unphysical large distance from the black hole. In other words, these low temperature phases may simply not exist for compact systems.}

\section{Discussion}
\label{windprop}
We use the best fit SED obtained by \cite{mar19} to compute the stability curves presented in Sec. \ref{stabcurvevol}. These SED have been computed in the JED-SAD framework and reproduce the spectral evolution of GX 339-4 during its 2010-2011 outburst. We show that the range of ionisation parameters consistent with Fe\ XXV and Fe XXVI ions is always in a thermally unstable part of the stability curves in the hard state. On the contrary, this range of ionisation parameters is always thermally stable in the soft state. Moreover, the observations of hot wind signatures in soft states indicate that the wind properties agree with this stable range of ionisation. Assuming that a wind is always present during the entire outburst, as we did in this work, we wonder then how the wind properties should evolve during the state transition while reaching (in the case of the hard-to-soft transition) or leaving (in the case of the soft-to-hard transition) the soft state ionisation conditions. Some interesting constraints need to be satisfied as discussed below. {As noticed in Sect. \ref{caveat} however, it is worth keeping in mind that the true evolution of the plasma would require detailed modeling of the disk wind physical properties and dynamics (e.g. \citealt{dan20,wat21}). While out of the scope of the present paper, they could significantly affect the following comments.}

\subsection{Changes in the disk wind properties during spectral transitions}

\subsubsection{From Hard-to-soft}
\label{HStrans}
As shown in Sect. \ref{tempevol}, only a few temperature domains are thermally stable during hard states (see Fig. \ref{figStabCurvTrans}b) either at very high temperature ($>3\times10^{7}$ K) or in localized region around $10^6$ K, $2\times10^5$ K or at low temperature $<3\times10^{4}$ K. As already said, the ionization range compatible with Fe\ XXV and Fe\ XXVI is itself thermally unstable. In the soft state however, a large part of the temperature domain is thermally stable including the ionization range compatible with Fe\ XXV and Fe\ XXVI. In order to  become detectable at the end of the Hard-to-Soft transition, the hot wind properties cannot be casual at the beginning. They have to evolve from the stable temperature domains of the hard states to the ``hot wind" domains of the soft states.\\

\begin{figure}[t]
\includegraphics[width=\columnwidth]{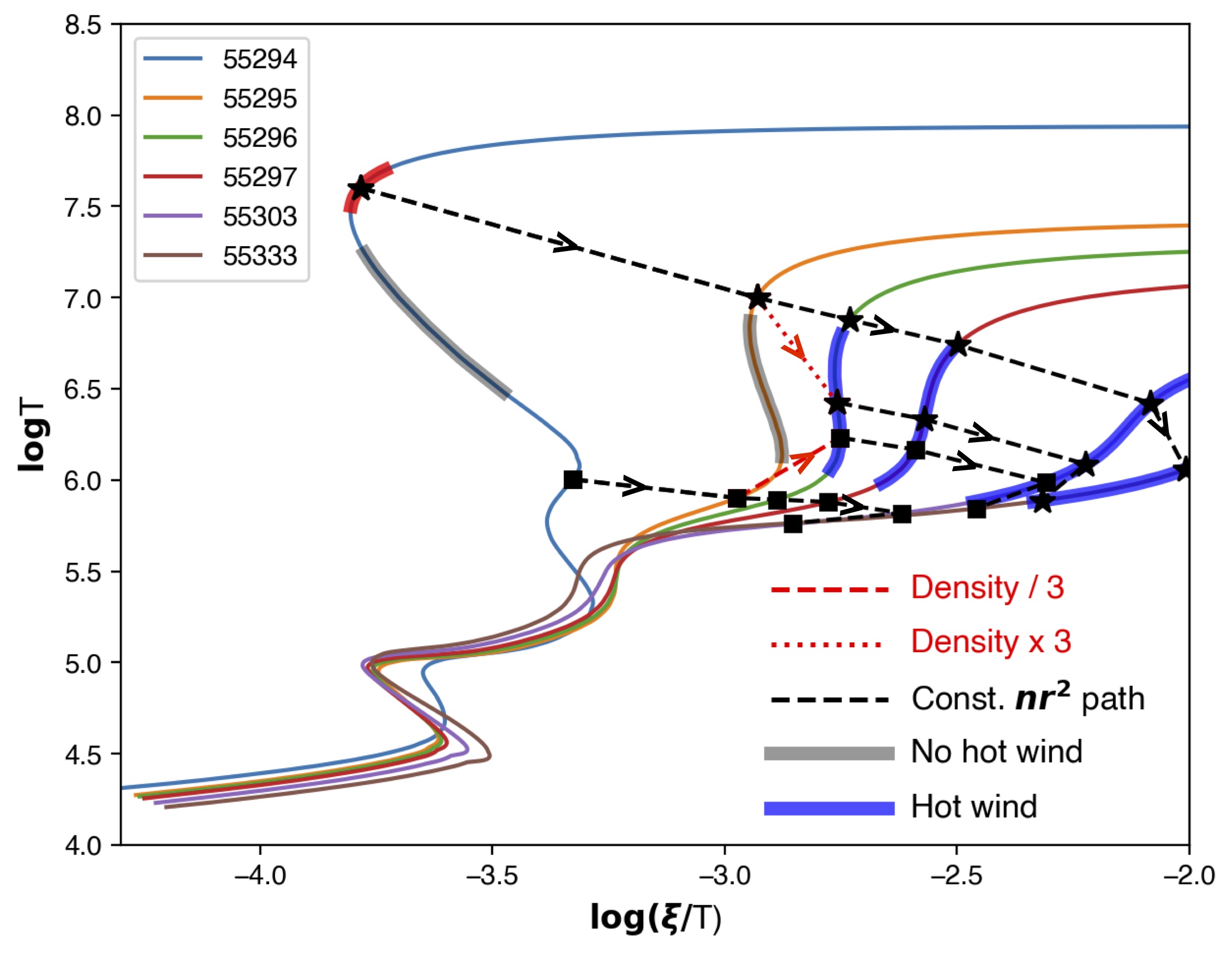} 
 \caption{The colored lines are the photoionisation stability curves during the Hard-to-Soft transition (see Fig. \ref{figStabCurvTrans} for the meaning of the blue and gray segments). The black dashed lines (squares/dashed line; star/dashed line) correspond to different paths with constant $nr^2$. 
 The red lines are the paths followed by the wind assuming an increase (for the red dotted path) or decrease (for the red dashed path) of the wind density by a factor 3 between MJD 55295 and MJD 55296. See Sect. \ref{HStrans} for the small temperature domain underlined in red on the stability curve of MJD 55294. The arrows indicate the time evolution.  }
\label{figcstenr2}
\end{figure}

Interestingly, this evolution is unlikely to occur at constant wind density distribution (i.e. constant $nr^2$) as shown in Fig. \ref{figcstenr2}. This figure is similar to Fig. \ref{figStabCurvTrans}d but we over-plotted two different paths (square/dashed line; star/dashed line) which correspond to two different wind evolutions at constant $nr^2$. The square/dashed line starts from the stable part of the temperature domain of the hard state around $10^6$ K (indicated by a square on the stability curve of MJD 55294). During the spectral evolution between MJD 55294 and MJD 55333, this path always stays on a stable part of the stability curves but it never crosses the ``hot wind'' domain of the soft states. In consequence, no signature of a hot wind will be visible in the soft states at the end of the transition. The same conclusion is obtained if the wind is initially (during MJD 55294) on one of the stable branches existing at lower temperature ($< 10^6$ K).  \\

The other path (star/dashed line) starts from the top hot stable branch ($T>3\times10^{7}$K) of the hard state, at the very beginning of this temperature domain (indicated by a black star on the stability curve of MJD 55294). During the spectral evolution of the system, this path crosses the stable ``hot wind" domains of the soft states only after MJD 55297. Actually only the constant $nr^2$ paths that start in the small temperature domain underlined in red on the stability curve of MJD 55294 will cross the ``hot wind'' domain of one of the soft states at the end of the transition. Indeed, a plasma at a temperature just below the red segment is unstable.  And for a plasma at a temperature larger than the red segment, constant $nr^2$ paths will never cross a ``hot wind" domain of the soft states and no wind will be detected at the end of the transition. This red segment corresponds however to a very limited range of temperature and ionisation parameter in the hard state. To reach the soft state, it appears more reasonable to assume a change in the wind physical characteristics during the transition (i.e. $nr^2$ does not stay constant). As an example, we over-plotted in red dashed lines the paths followed by the wind if its density increases (red star/dashed line) or decreases (red square/dashed line) by a factor 3 between MJD 55295 and MJD 55296. In both cases, the change in density modifies significantly the paths compared to the constant $nr^2$ cases and they now cross the ``hot wind'' domains during almost all the soft state. Although empirical, this strongly supports the need for a change in the wind properties during the Hard-to-Soft transition in order to have detectable Fe\ XXV and Fe\ XXVI lines in the soft states.
Actually the minimal increase of density to reach a stable "hot wind" domain when coming from the black star/dashed line path is of about 30\% while the minimal decrease of density when coming from the black square/dashed line path is of about a factor 2. 

\subsubsection{From Soft-to-Hard}
\label{SHtrans}
\begin{figure}[t]
\includegraphics[width=\columnwidth]{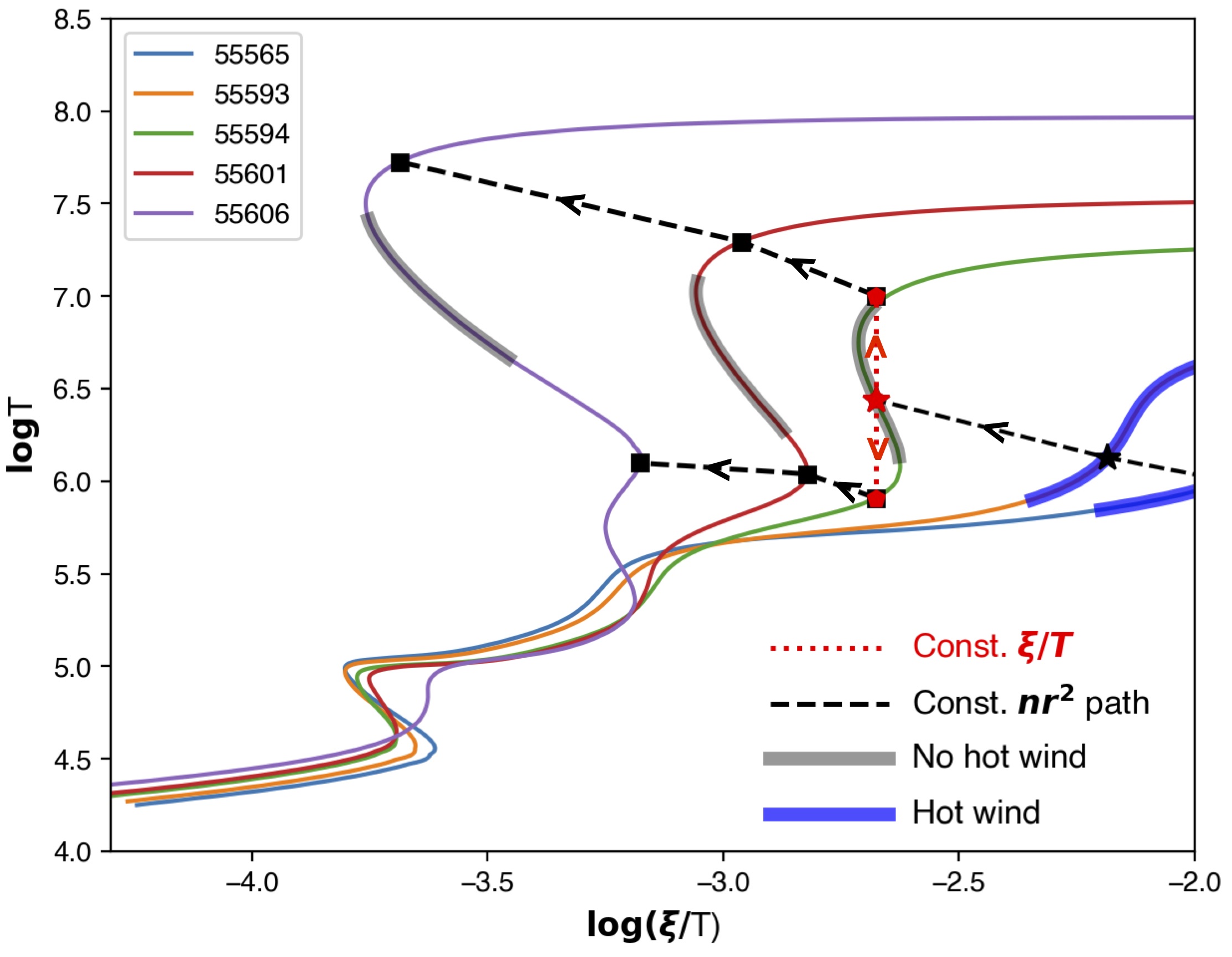} 
 \caption{The colored lines are the photoionisation stability curves during the Soft-to-Hard transition (see Fig. \ref{figStabCurvTrans} for the meaning of the blue and gray segments). The black stars/dashed line correspond to a path of constant $nr^2$ from MJD 55565 until MJD 55594. The vertical red dotted lines correspond to isobaric (constant $\xi/T$) evolutions of the wind. The black square/dashed lines correspond to path of constant $nr^2$ starting from MJD 55594. The arrows indicate the time evolution.}
\label{figcstenr2b}
\end{figure}

We can repeat a similar procedure to follow the evolution of the wind thermal state during the transition from the soft to the hard state. This is illustrated now in Fig. \ref{figcstenr2b}. Let us assume that at the beginning of the transition the wind is in the conditions where Fe\ XXV and Fe\ XXVI lines are detectable (black star on the MJD 55593 stability curve). Then a path at constant $nr^2$ (black dashed line) reaches the unstable temperature domain of the stability curve of MJD 55594 (black star on the MJD 55594 stability curve). 
Actually, looking at the shape of the stability curves, this evolution does not depend strongly on the starting point in the ``hot wind'' domain of the soft states (MJD $<$ 55594), and, as long as $nr^2$ stays constant, the wind will inevitably reach an unstable state when the X-ray source enters in the hard state at MJD 55594.\\ 

The evolution of the wind after MJD 55294 is thus quite unknown. The thermal instability forces the wind plasma to leave the thermally unstable temperature domain and to reach the other stable branches of the stability curve. This necessarily requires a change of the wind temperature and density. Locally, as soon as the plasma reaches a new (stable) temperature, the new ionization equilibrium is expected to settle in almost immediately \citep{gon07}. Of course, the complete migration of the gas from a thermally unstable to a thermally stable state, which is itself a mix of hot and cold phases in interaction, will take some time. This is expected to occur on a time scale $t_{migr}$ of the order of $\Delta H/c_s$ where $\Delta H$ is the thickness of the thermally unstable region and $c_s$ is the sound velocity in the cold phase (see discussion in \citealt{gon07}). This translates to:
\begin{equation}
t_{migr}\lesssim 2\times 10^4 \frac{N_{23}}{T_5^{1/2}n_{12}} \mbox{ (in units of s)}
\label{tdyn}
\end{equation}
where $N_{23}$ is the column density in units of $10^{23}$ cm$^{-2}$ , $n_{12}$ is the
hydrogen density in units of $10^{12}$ cm$^{-3}$ and $T_5$ is the temperature of
the cold gas in units of $10^5$ K . 
Assuming a maximal extension of the disk wind of the order of the binary separation in GX 339-4, i.e. $\sim 10^{12}$ cm \citep{zdz04b,zdz19}, and an optically thin wind, $t_{migr}$ is shorter than a day. So the migration can easily occur between two {\it RXTE} pointings (generally separated by 1 day).\\

The vertical dotted segments in Fig. \ref{figcstenr2b} indicate the cases where the evolution of the gas after MJD 55594 occurs at constant pressure (isobaric evolution i.e. a constant $\xi/T$) and fast enough to assume that the average distance of the gas to the X-ray source would not change. In this case, the wind plasma can transition to two different stable solutions, one hotter and the other colder, indicated by the upper and lower red squares respectively on the MJD 55594 stability curve. It is also possible that the wind  evolves into a mix of these two different plasma phases where colder (and denser) structures are embedded in a hotter (less dense) gas. This situation would be very similar to the one discussed by \cite{bia17} in the case of the wind evolution observed in the neutron star AX J1745.6-2901 between its soft and hard state.
After MJD 55594, each of the two possible wind phases will keep on evolving depending on the spectral evolution of the X-ray source. As examples, we over-plotted  in Fig. \ref{figcstenr2b} in red dashed lines the two possible paths where $nr^2$ is assumed to be constant.

\subsubsection{A very favorable configuration for disk wind detection}
Interestingly, during the soft-to-hard and hard-to-soft transitions, the shape of the stability curve becomes almost vertical in the ionization range compatible with Fe\ XXV and Fe\ XXVI ions (see e.g. the stability curve of MJD 55296 in Fig. \ref{figStabCurvTrans}h). This corresponds to a situation where pressure equilibrium is satisfied in this ionisation domain and this is a quite favorable mechanical configuration for the wind. This could explain why we generally observe the strongest wind absorption features in the middle of the transition states (see \citealt{pon12}).

\subsection{Disk irradiation as a potential driver of the change of the disk wind properties during the hard-to-soft transition}
\label{disksurfheat}
%
%
%
As shown in Sect. \ref{HStrans}, if we want to observe a hot wind at the end of the hard-to-soft transition, an increase (or a decrease depending on the wind phase at the beginning of the transition) of the wind density seems to be needed to make the wind evolving into the stable ionization range for Fe\ XXV and Fe\ XXVI ions.

It is hard to estimate the timescale for such evolution given the lack of long (several days) and continuous monitoring of a hard-to-soft state transition of an XrB. 
But the fact that absorption features are regularly observed in soft states suggests that the wind properties should evolve rapidly (on $\sim$ days) after its entry in the soft state. This is somewhat problematic if the wind density scales only with the local disk density. Since the latter scales with the disk accretion rate, the wind density would be expected to increase and then decrease during the outbursting cycle. However,  the disk wind region is estimated to be located quite far away, beyond $10^4 R_g$. There is therefore no reason why the slow time evolution of the far-out wind driving zone should always be contemporaneous with the rapid spectral changes of the SED emitted from the innermost disk regions. Irradiation of the outer disk by the inner X-ray emission seems the most natural process to link these two very distant regions on a very short timescale. Such a radiative feedback of the inner regions on the outer disk wind properties is plausible, since the deposition of any additional power  $Q$  at the disk surface leads to the enhancement of mass loss, both in magnetically-driven (e.g. \citealt{cas00b})  and thermally-driven (e.g. \citealt{hig15}) outflows.\\ 

The amount of X-ray luminosity incident on the disk at a radius $r=R/R_g$ in a ring of thickness $\Delta r$ will depend on the exact geometry. Estimates from simple radial profile of the disk aspect ratio provide a too low X-ray irradiation which does not fit with the required disk accretion transport to explain the XrB outburst light curve durations (e.g. \citealt{dub01,tet18b,tet20}). Following \cite{dub01}, the intercepted and reprocessed X-ray irradiation (which corresponds to the heating power deposit $Q$ we are looking for) by a disk ring can be simply expressed as:
\begin{equation}
Q \simeq C_{irr} \frac{L_X}{4\pi r^2}2\pi r dr=C_{irr} L_X\frac{\Delta r}{2 r}
\label{eqQ}
\end{equation}
with  $C_{irr}$ a parameter which encapsulates our ignorance concerning the irradiation geometry, the X-ray albedo and X-ray irradiating spectral shape. $C_{irr}$ is estimated to be  $\simeq 0.05$ to explain the XrB outburst light curve durations (see \citealt{tet18} and references therein). Actually, a (weak) radial dependence of $C_{irr}$ is expected due to the radial variation of the disk aspect ratio (see \citealt{dub19} and references therein). The scattering of the X-rays by the disk wind itself can also have some impact on the disk irradiation \citep{dub19,kim19}. But for the present discussion we will assume $C_{irr}$ to be constant equal to 0.05. Equation \ref{eqQ} then shows that the heating through irradiation is directly related to the X-ray luminosity. The thermal timescale on which the disk matter is heated is given by (e.g. \citealt{gon07}):
\begin{equation}
t_{th}\sim\frac{T_5}{n_{12}\Lambda_{23}} sec.
\end{equation}
where, like in Eq. \ref{tdyn}, $n_{12}$ is the density expressed in $10^{12}$ cm$^{-3}$, $T_5$ is  the  temperature in $10^5$ K and $\Lambda_{23}$ is the cooling  function in 10$^{-23}$ erg cm$^{3}$ s$^{-1}$. It is thus reasonable to expect thermalization time scales of seconds to minutes. The travel time for the X-ray photons produced at a few $R_g$ from the black hole to reach the outer disk ($R>10^4 R_g$) is of the same order of magnitude. Finally the density adjustment of the wind will follow the thermal adjustment of the underlying disk on, typically, the local keplerian time scale which is less than a day even at a disk radius of $10^5 R_g$ (assuming a 10 solar mass black hole). Overall the entire process (irradiation and wind density adjustment) is expected to occur in less than a day.
 \\ 

We have reported the X-ray luminosity $L_X$ computed from 
the SED between 1 and 100 keV in Tab. \ref{param}. 
The luminosity $L_X$  increases rapidly at the beginning of the transition with an increase of a factor $\sim$1.5-2 between MJD 55294 and MJD 55303 while the accretion rate stays roughly constant (see Tab. \ref{param}). Our qualitative estimates then suggest an increase of the disk surface heating through irradiation by the inner X-ray source during the hard-to-soft transition and, consequently, a potential increase of the disk wind mass loading and wind density in a short (less than a day) timescale. Thus, looking at Fig. \ref{figcstenr2}, our result supports the scenario with an increase of the wind density during the transition. A more quantitative  impact on the wind density would require however a precise modeling of the thermal balance of the disk outflowing disk surface layers which is out of the scope of the present paper.


\section{Concluding remarks}
\label{conclusion}
We studied the expected evolution of the thermal properties of a hot disk wind (e.g. characterized by the presence of absorption lines of Fe\ XXV and Fe\ XXVI ions) in a low-mass X-ray binary assuming it is present during the entire outburst (e.g. \citealt{san20}). For that purpose we have used the best fit SEDs obtained by \cite{mar19} to reproduce the {\it RXTE/PCA} data of GX 339-4 during the 2010-2011 outburst of the source. These broad-band SEDs have been obtained in the Jet Emitting Disk framework described in \cite{fer06a,mar18b,mar18a}. For each of the 274 SEDs, we computed the corresponding stability curve using the CLOUDY software. We were then able to follow the evolution of these stability curves all along the outburst. Our main conclusions are:
\begin{itemize}
\item all hard-state observations are characterized by stability curves where the ionisation domain compatible with  Fe\ XXV and Fe\ XXVI ions is thermally unstable. On the contrary, in all the soft state observations this ionisation domain is thermally stable.
\item during the spectral state transitions, the evolution of the stability curve from stable to unstable conditions (in the soft-to-hard transition) and reversely (in the hard-to-soft transition) can occur rapidly (day timescale) due to the change of the illuminating ionizing SED. 
\item during the hard-to-soft transition, the common observations of hot wind signatures in the soft states and their absence in the hard states suggests a rapid evolution of the disk wind properties, most likely a significant increase (or decrease depending on the ionisation state of the wind at the beginning of the transition) of its density. Moreover, this evolution must occur quasi simultaneously with the change of the spectral state of the inner X-ray source located $10^4$ to $10^5$ $R_g$ away from the wind production site. 
\item heat deposit at the disk surface through X-ray irradiation by the inner X-ray source could drive the required increase of the disk wind density in a short timescale (less than a day).
\item during the soft-to-hard transition, if a hot wind is present in the soft states, the hardening of the X-ray emission puts unavoidably the hot wind into an unstable part of the stability curve in the hard states. Then the wind plasma necessarily moves to a hotter or a colder (or a mix of the two) phase where Fe\ XXV and Fe\ XXVI ions cannot exist and, in consequence, cannot be detected. 
\end{itemize}
{{Note that the first item is also consistent with our initial assumption that hot winds are always launched from the outskirts of the accretion disk, regardless of the spectral state of the innermost region. But they are only observable in soft states during which the required thermal stability occurs. }}

{{Recent numerical simulations (e.g. \citealt{zhu18,jac19} and references therein) show that disk winds are expected from low magnetized turbulent accretion disks threaded by a large-scale magnetic field, {which would correspond to our SAD}. For high magnetization however, {which would correspond to our JED}, powerful jets are rather produced (e.g. \citealt{lis20}). Our results thus provide a case for the presence, in XRBs, of a large scale magnetic field throughout the entire accretion disks and whose magnetization would evolve from high values in the inner regions {(i.e. \cite{lis20} configuration)}, to much lower values in the outer regions, where a disk wind would be launched ({i.e  \cite{zhu18} or \cite{jac19} configuration)}. In recent 3D simulations \citep{jac20}, such a radial stratification of the magnetization is expected as the outcome of a global readjustment of the disk (see also \citealt{sce19,sce20}). If true, this could have a significant impact on the secular evolution of the accretion flow.}}

%

\label{discussion}

\begin{acknowledgements}
Part of this work has been done thanks to the financial supports from CNES and the French PNHE. SBi acknowledges financial support from the Italian Space Agency under grant ASI-INAF 2017-14-H.O. GP acknowledges funding from the European Research Council (ERC) under the European Union's Horizon 2020 research and innovation program (grant agreement No. [865637]).

\end{acknowledgements}


\begin{appendix}

\section{{{The JED-SAD framework}}}
\label{app1}
{{The SED used in this study have been computed in the JED-SAD framework described in details in \cite{mar18a,mar18b} and \cite{mar19}. While the reader is referred to these papers for more details, we briefly recall here the main characteristics of this model. JED solutions assume the existence of a large scale vertical magnetic field $B_z$ threading the disk. The JED magnetization $\mu=B_z/(\mu_0P_{tot})$, where $P_{tot}$ includes the plasma and radiation pressures, is in the range [0.1-1]. In these conditions, in a JED, part of the accreted matter is ejected and the accretion rate is a function of the radius i.e. $\dot{M}(R)\propto R^p$ with $p$ the ejection parameter\footnote{The ejection parameter is labeled $\xi$ in \cite{mar19}. It has been changed to $p$ to not be confused with the ionisation parameter of the wind.}. In contrast with the ADIOS model \citep{bla99,beg12} 
the  ejection efficiency $p$ is not a free parameter in the JED model, but is a consequence of the disk properties (magnetization, turbulence). Its value results from the resolution of the full set of MHD equations governing the interdependent accretion-ejection physics in the presence of a large scale magnetic field (see \citealt{fer02,fer08} for reviews).}}\\  

{{The SEDs of a JED-SAD configuration combine the emission of the JED, present between $R_{ISCO}$ and a transition radius $R_{tr}$, and the emission of the non-ejecting SAD extending beyond $R_{tr}$. The JED thermal structure is computed taking into account the local turbulent heating, coulomb heat exchange (i.e. a two-temperature plasma), heat advection and cooling using a bridge formula for optically thick and thin plasmas. The optically thin cooling processes are bremsstrahlung, synchrotron and local and non local (on the soft photons emitted by the outer SAD) Compton scattering  processes.  The SED depends mainly on {{two}} parameters: the transition radius $R_{tr}$ and the inner disk accretion rate $\dot{M}(R_{ISCO})$ (simply noted $\dot{M}$ hereafter) {{at $R_{ISCO}$}}. The global spectrum emitted by a JED-SAD disk configuration characterizes by ($R_{tr}$, $\dot{M}$) is  self-consistently computed along with the disk thermal and dynamical states. {{It appears that the JED can play the role of the hot corona needed to explain the hard X-ray emission in XrB. Its emission dominates the high energy part of the SED while the multi-color disk emission from the SAD dominates at low energy \citep{mar18b}}}. Note that for $R_{tr}=R_{ISCO}$, there is no JED. The entire accretion flow, then, is a SAD and the source is in a soft state. On the contrary, large $R_{tr}$ ($\gg R_{ISCO}$) characterize hard states, while $R_{tr}$ of the order of a few $R_{ISCO}$ are typical of (either hard or soft) intermediate states. }}

\begin{figure}[h]
\begin{center}
\includegraphics[width=\columnwidth]{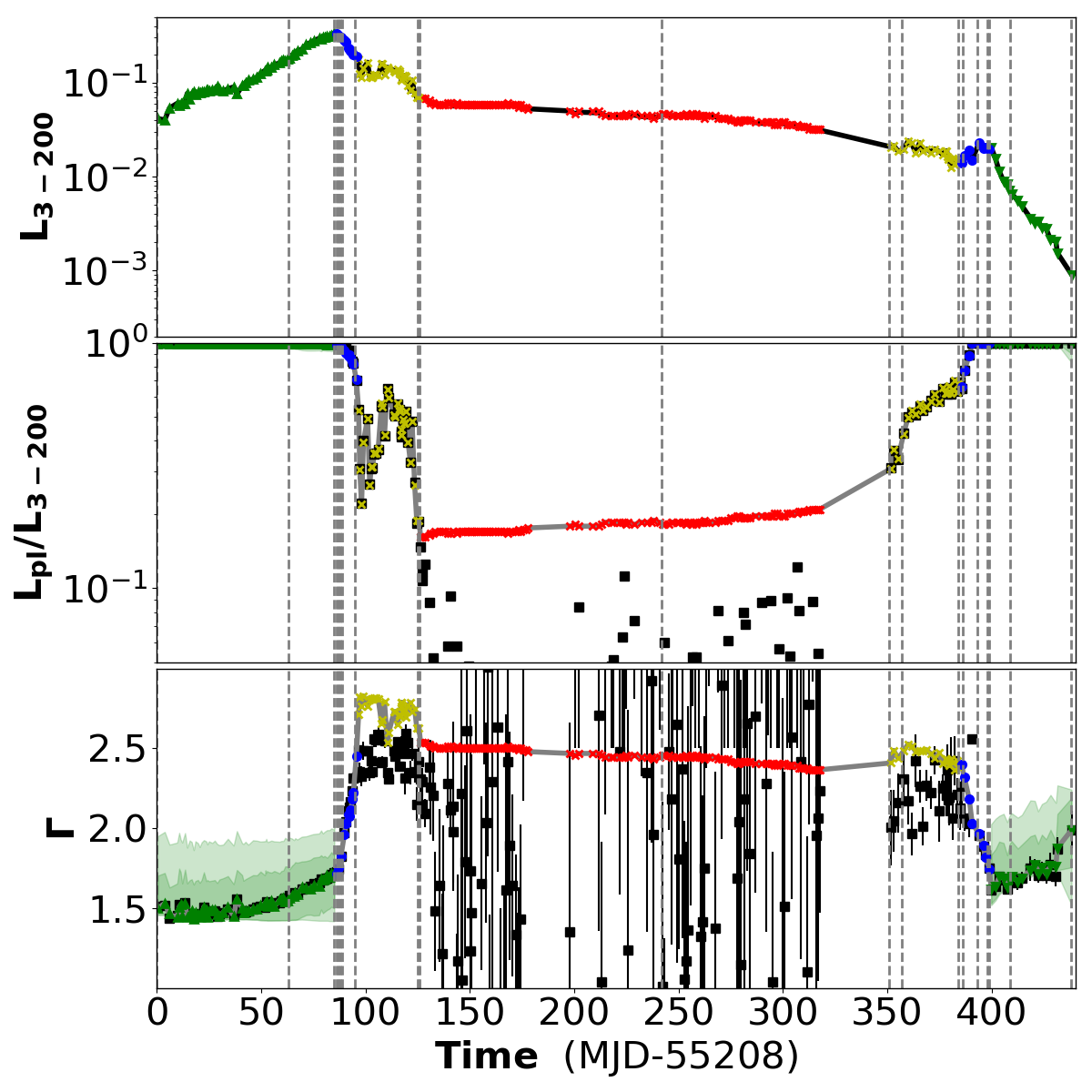}
 \caption{{{Results of the fitting procedure of \cite{mar19} applied to the 2010-2011 outburst of GX 339-4. The black markers are fits taken from \cite{cla16} reported with their error bars when reliable, while color lines display the best fit values obtained with the JED-SAD SED: green, blue, yellow, and red for hard, hard-intermediate, soft-intermediate, and soft states, respectively. From top to bottom: 3-200 keV total luminosity, $L_{3-200}$ (in Eddington units), the power-law luminosity fraction $L_{pl}/L_{3-200}$, defined as the ratio of the power-law flux to the total flux in the 3-200 keV range, and the power-law index $\Gamma$. The transparent colored areas correspond to the confidence intervals of 5\% and 10\% error margin (see  \citealt{mar19}). The vertical dashed line correspond to the MJD reported in Tab. \ref{param}.}}}
\label{SEDex}
\end{center}
\end{figure}

{{This model has been applied with success to the {\it RXTE/PCA} data of the 2010-2011 outburst of GX 339-4. In this case, it is assumed for $R_{ISCO}$ a black-hole mass of 5.8 solar masses\footnote{Multiple studies have been performed to constrain the black hole mass in GX 339-4 and have led to different estimations (e.g., \citealt{hyn03,mun08,par16,hei17}). For simplification, and since there is no consensus about GX 339-4 mass, we chose it to be the central value.} and a black hole spin $a$=0.94 (see references in \citealt{mar19}). The distance of GX 339-4 is also assumed to be 8 kpc \citep{hyn04, zdz04b,par16}. These results are discussed in details in \cite{mar19} (the 3 other outbursts of this source during the {\it RXTE} era have been analyzed in \citealt{mar20}). We have reported the comparison between the observations and the model in Fig. \ref{SEDex}. The observed values of the 3-200 keV total luminosity, the power-law luminosity fraction $L_{pl}/L_{3-200}$ (defined as the ratio of the power-law flux to the total flux in the 3-200 keV range) and the power-law index $\Gamma$ obtained by \cite{cla16} are reported with black markers at the top, middle and bottom panels respectively. The corresponding values produced by the best fit SED obtained by \cite{mar19} are over-plotted with colored markers. The MJD of Tab. \ref{param} are also indicated with dashed vertical lines. The best fit SEDs reproduce the observations very well except in the soft state for what concerns the power law component (middle and bottom panels, red markers). In the soft states however, the statistics at high energy ($>$10 keV) of the RXTE/PCA data is too low to provide good constraints on this power law and the comparison is not relevant.}}

\begin{figure*}[h]
\begin{tabular}{cc}
\includegraphics[width=0.5\textwidth]{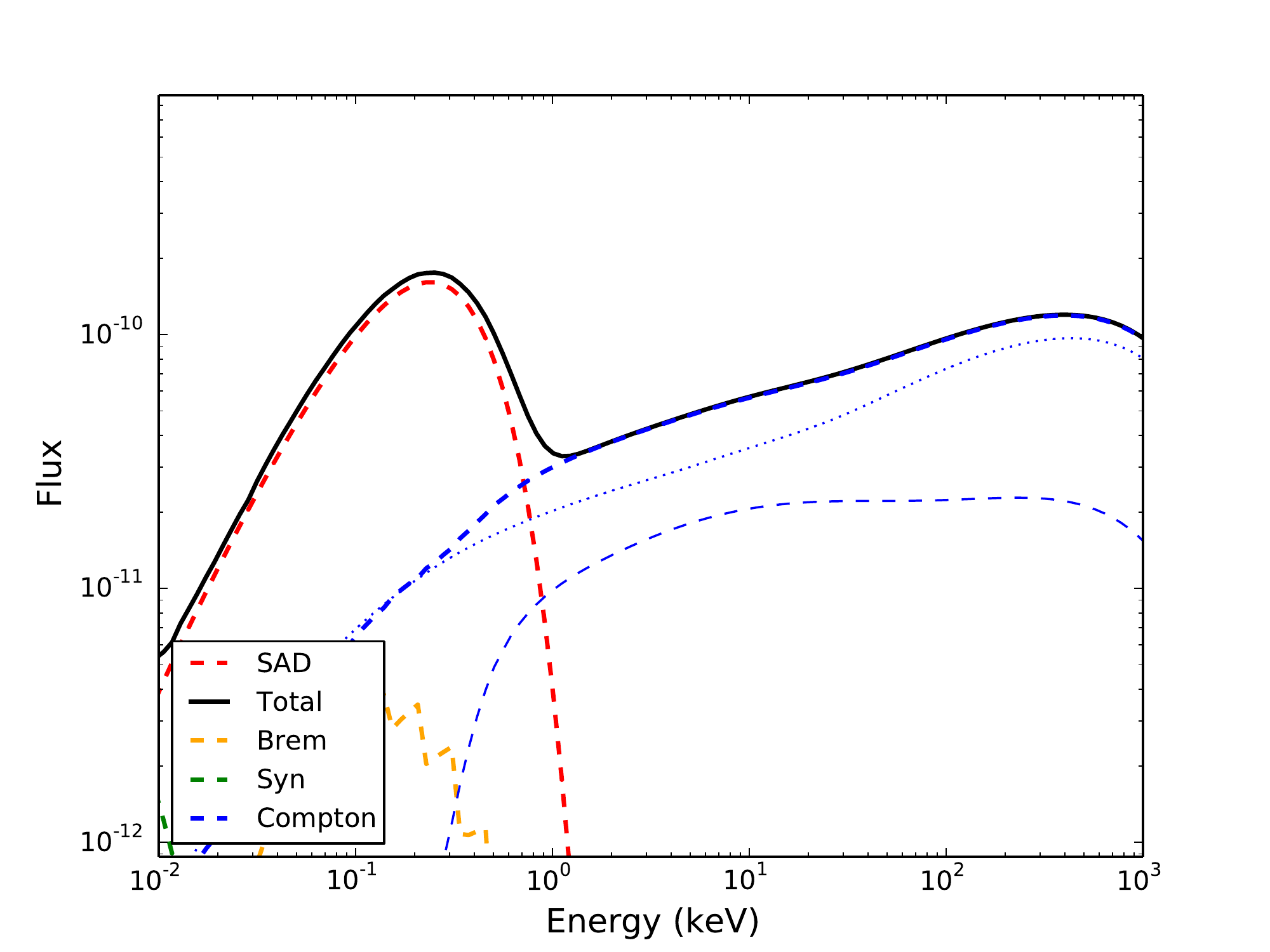} &
\includegraphics[width=0.5\textwidth]{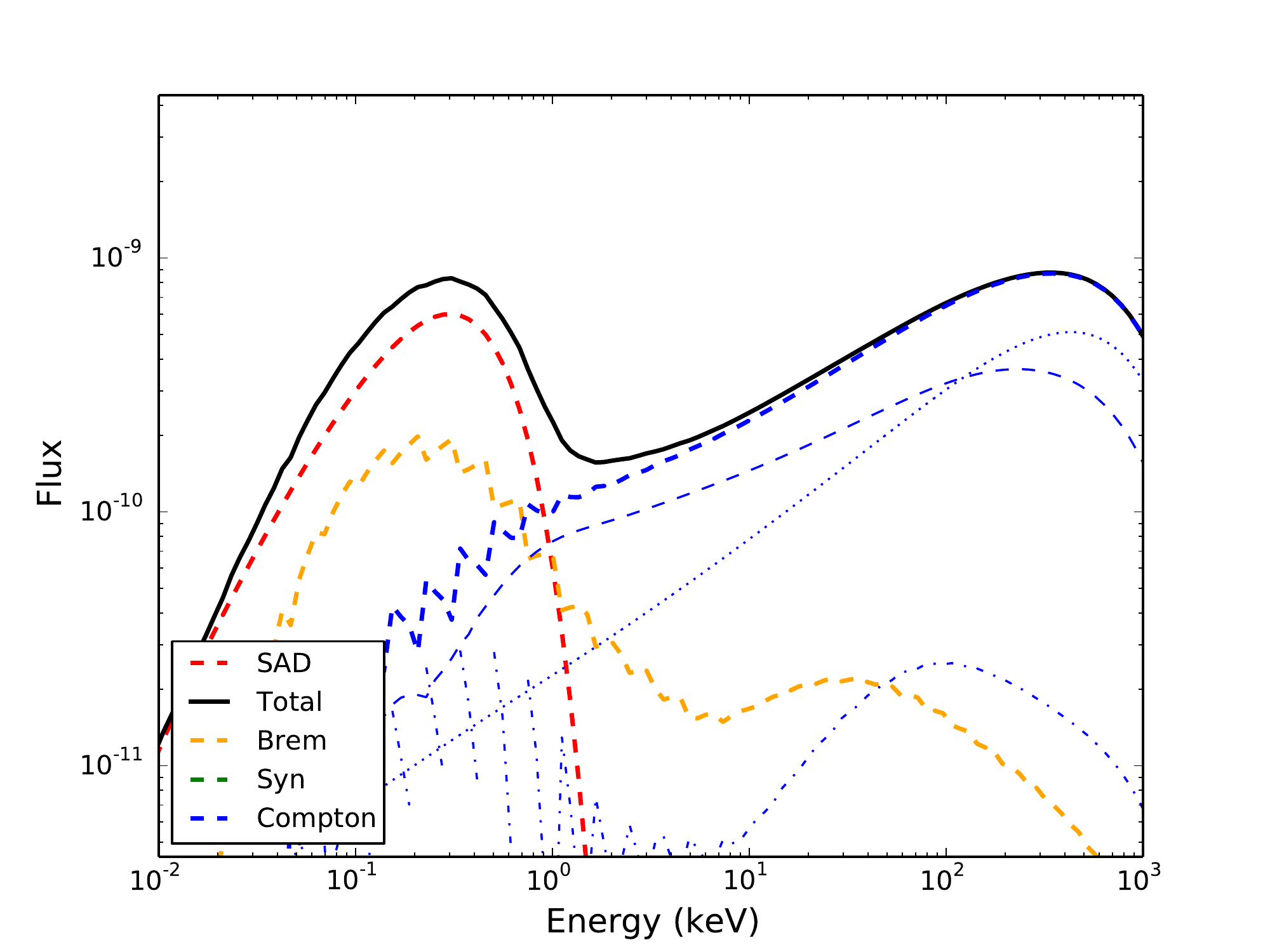}\\
\includegraphics[width=0.5\textwidth]{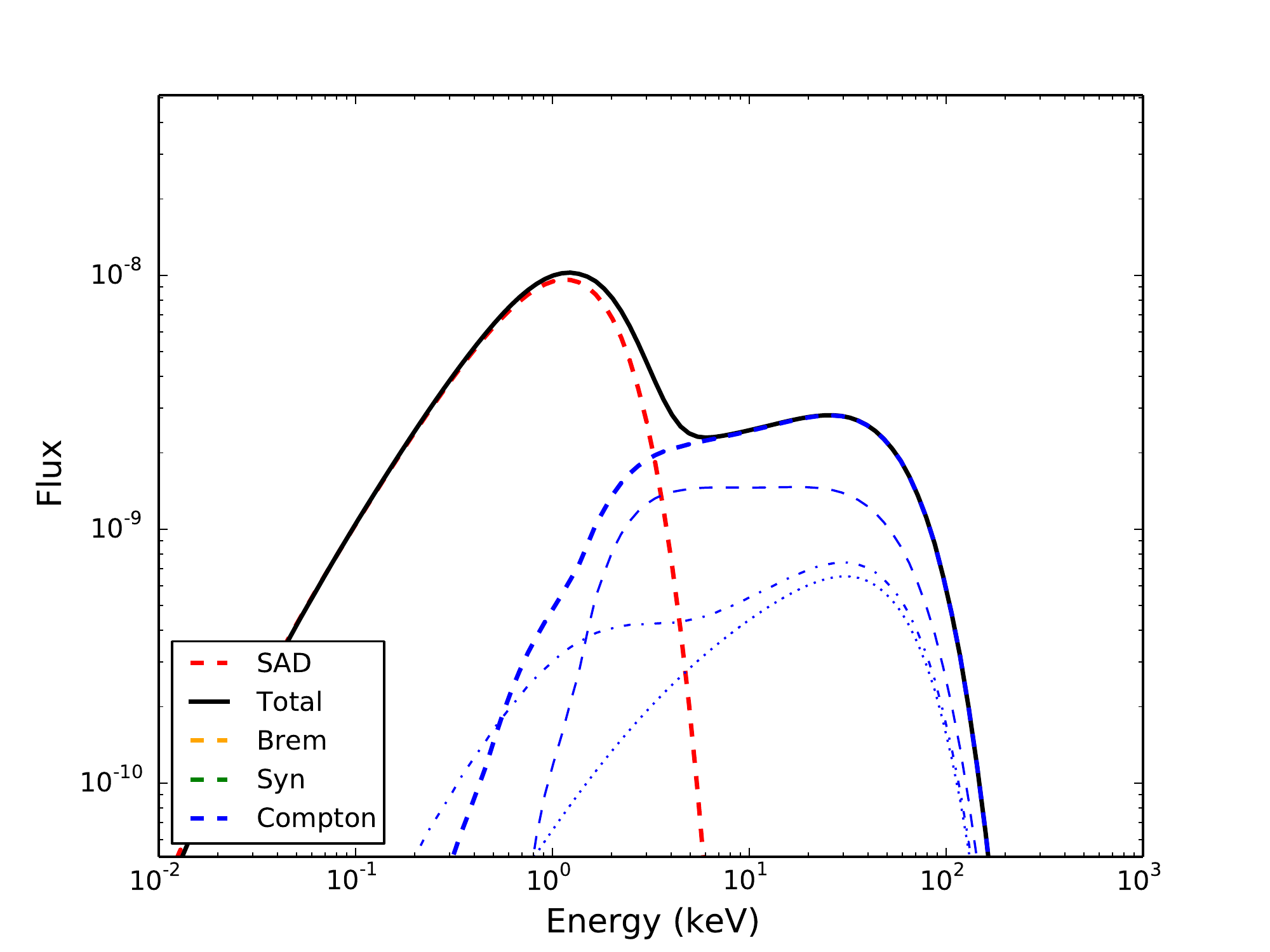} &
\end{tabular}
\caption{{JED-SAD SED for MJD 55617 (top left), 55208 (top right) and 55297 (bottom left). The different radiative processes (bremsstrahlung, synchrotron and Compton) are plotted in colored thick dashed lines (see the legends). We have also separated the Compton component in external Compton (blue thin dashed line), comptonized bremsstrahlung (blue thin dot-dashed line) and comptonized synchrotron (blue thin dotted line).
\label{SEDcompton}.}}
\end{figure*}

{We have also reported in Fig \ref{SEDcompton} the SED for MJD 55617, 55208 and 55297 detailing the importance of the different radiative processes (bremsstrahlung, synchrotron and Compton)}.
%

\section{{{Effect of density on the stability curve shape}}}
\label{effectdens}
\begin{figure}
\begin{tabular}{c}
\includegraphics[width=\columnwidth]{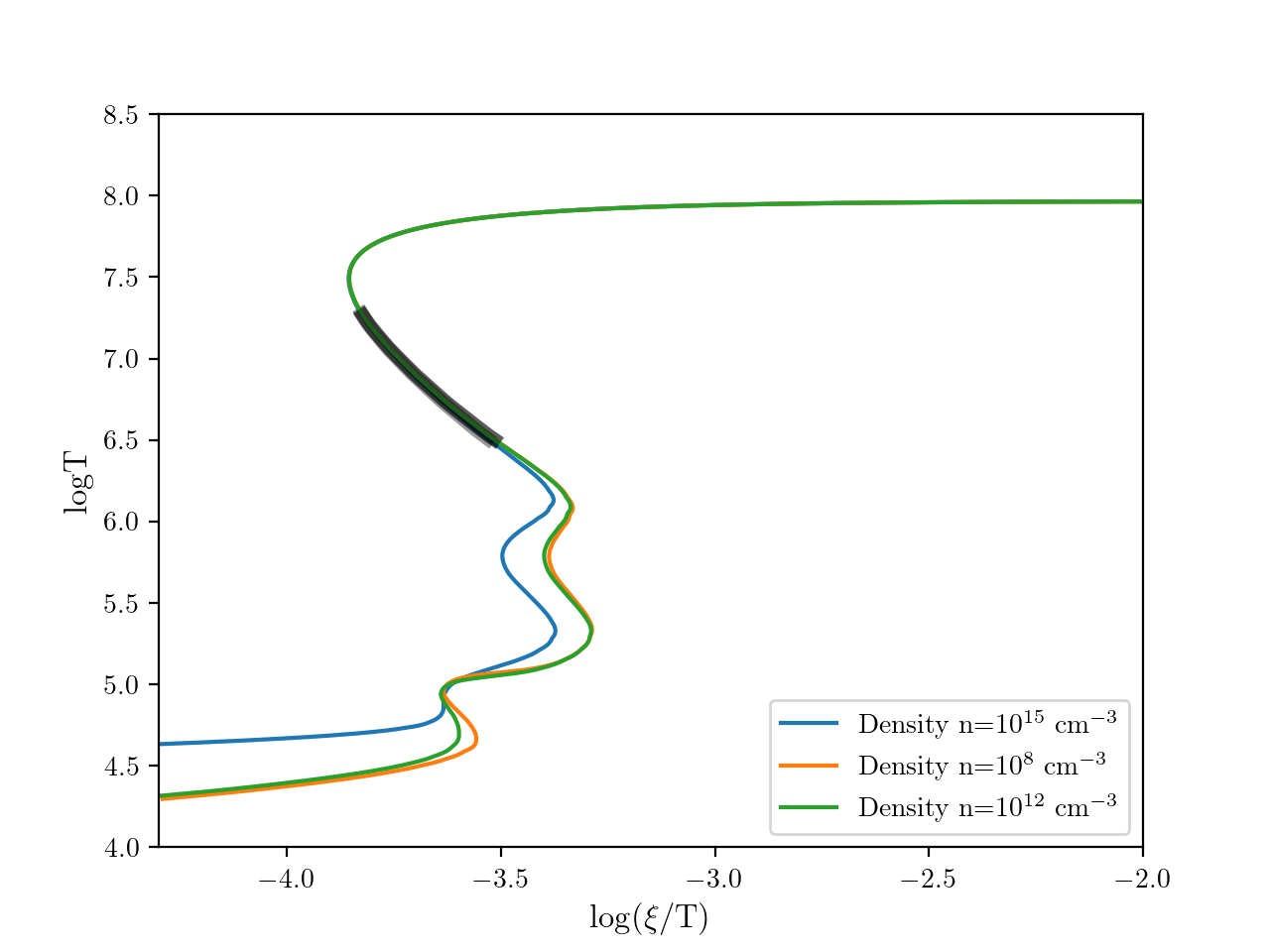}\\
\includegraphics[width=\columnwidth]{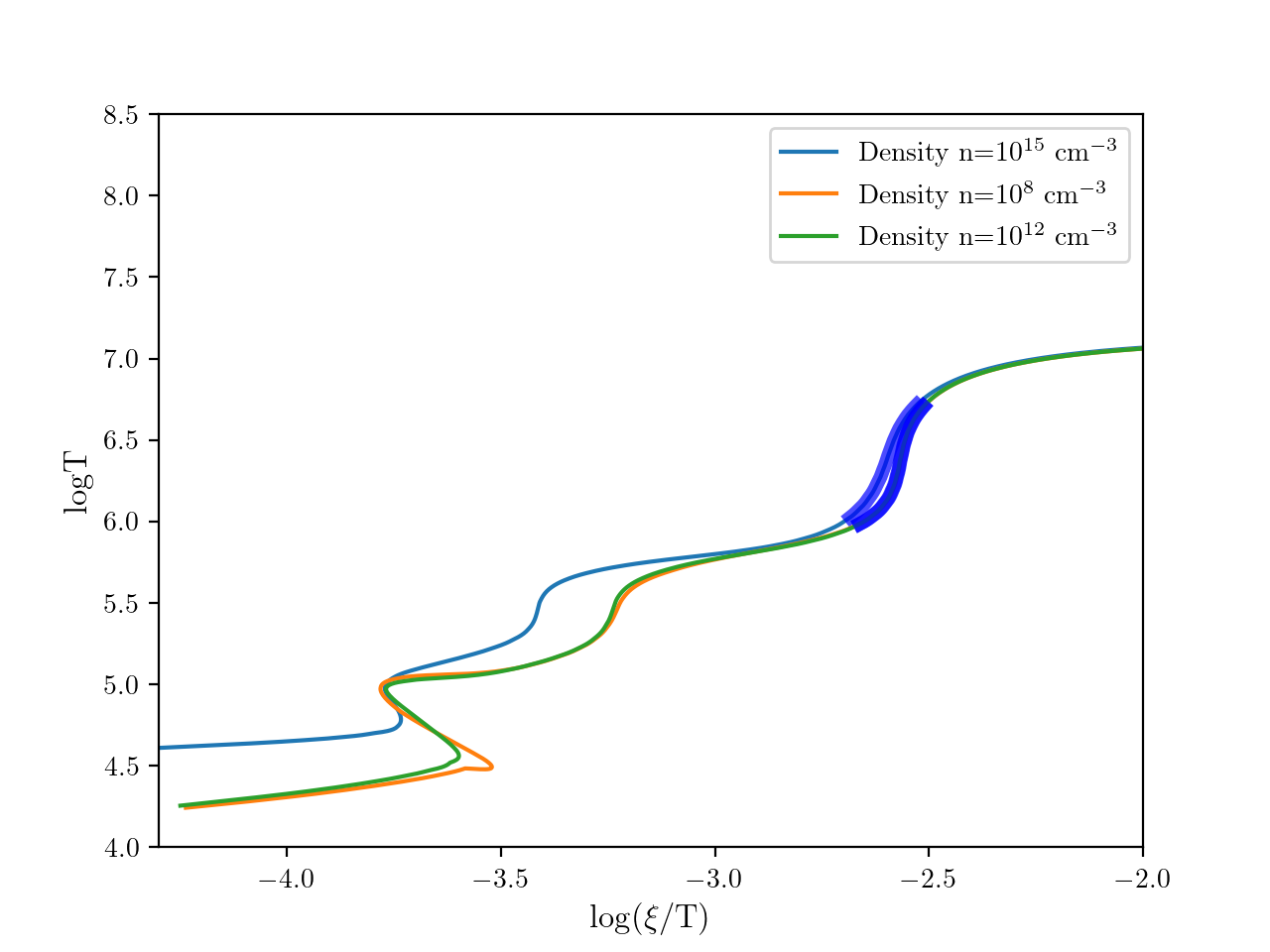}
\end{tabular}
 \caption{{{Stability curves computed with Cloudy for MJD 55293 (top) and MJD 55297 (bottom) for different values of the plasma density varying between $10^8$ and $10^{15}$ cm$^{-3}$.}}}
\label{effectdensity}
\end{figure}

{{We have reported in Fig. \ref{effectdensity} the stability curves computed with Cloudy for MJD 55293 (top) and MJD 55297 (bottom) for different values of the plasma density varying between $10^8$ and $10^{15}$ cm$^{-3}$. If  differences at low $\xi$ are visible, although not large enough to qualitatively change the general behavior of the curves, at high $\xi$ the stability curves are almost independent on the density. The reason is that Compton scattering dominates there and both heating and cooling terms scale with density in the same way.}}  

\section{Ion fraction}
\begin{figure}[h]
\begin{center}
\includegraphics[width=\columnwidth]{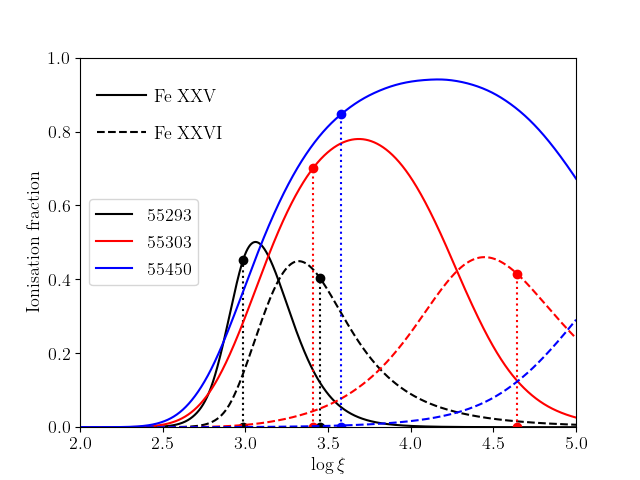}
 \caption{Fe\ XXV (solid lines) and Fe\ XXVI (dashed lines) ion fractions for three MJD of the 2010-2011 outburst of GX 339-4. They have been computed by using the SED of \cite{mar19}. The vertical dotted lines define, for each curve, the range in $\log\xi$ used to create the highlighted (blue or gray) areas on the stability curves presented in this paper. The lower limit in $\log\xi$ for each range corresponds  to an ionic fraction equal to 90 per cent of the ionic fraction peak of Fe\ XXV, while the upper limit corresponds to an ionic fraction equal to 90 per cent of the ionic fraction peak of Fe\ XXVI. If this upper limit is larger than 5 (like for MJD 55450), it is put equal to 5. }
\label{ionfraction}
\end{center}
\end{figure}

We have reported in Fig. \ref{ionfraction} three examples of ion fractions for three MJD of the 2010-2011 outburst of GX 339-4.

\end{appendix}

\end{document}